\documentclass[twocolumn,tighten,times]{aastex62}
\usepackage{amsmath}
\submitjournal{ApJ}
\shorttitle{Spiral Arm Kinematics in M51}
\shortauthors{Pineda et al.}

\begin{document}
\title{A SOFIA Survey of [C\,\sc ii] in the galaxy M51 II. [C\,{\sc
    ii}] and CO kinematics across spiral arms}

\author[0000-0001-8898-2800]{Jorge L. Pineda}
\affiliation{Jet Propulsion Laboratory, California Institute of Technology, 4800 Oak Grove Drive, Pasadena, CA 91109-8099, USA} 
\author[0000-0001-7658-4397]{J\"urgen Stutzki}
\affiliation{KOSMA, I. Physikalisches Institut, Universit\"at zu K\"oln, Z\"ulpicher Stra\ss e 77, 50937\\ K\"oln, Germany} 
\author[0000-0002-2064-7691]{Christof Buchbender}
\affiliation{KOSMA, I. Physikalisches Institut, Universit\"at zu K\"oln, Z\"ulpicher Stra\ss e 77, 50937\\ K\"oln, Germany} 
\author[0000-0002-8762-7863]{Jin Koda}
\affiliation{Department of Physics and Astronomy, Stony Brook University, Stony Brook, NY 11794-3800, USA} 
\author[0000-0003-2649-3707]{Christian Fischer}
\affiliation{Deutsches SOFIA Institut, Pfaffenwaldring 29, 70569 Stuttgart, Germany}
\author[0000-0002-6622-8396]{Paul F. Goldsmith}
\affiliation{Jet Propulsion Laboratory, California Institute of Technology, 4800 Oak Grove Drive,\\ Pasadena, CA 91109-8099, USA} 
\author[0000-0001-6708-1317]{Simon C. O. Glover}
\affiliation{Universit\"at Heidelberg, Zentrum f\"ur Astronomie, Albert-Ueberle-Str. 2, D-69120 Heidelberg, Germany}
\author[0000-0002-0560-3172]{Ralf S. Klessen}
\affiliation{
Universit{\"a}t Heidelberg, Interdisziplin{\"a}res Zentrum f{\"u}r Wissenschaftliches Rechnen, Im Neuenheimer Feld 205,  69120 Heidelberg, Germany}
\author[0000-0002-4908-4925]{Carsten Kramer}
\affiliation{Institut de Radioastronomie Millim\'{e}trique (IRAM), 300 rue de la
  Piscine, 38406 Saint Martin d'H\`{e}res, France}
\author[0000-0003-1766-6303]{Bhaswati Mookerjea}
\affiliation{Tata Institute of Fundamental Research, Homi Bhabha Road, Mumbai, 400005, India}
\author[0000-0002-0820-1814]{Rowan Smith}
\affiliation{Jodrell Bank Centre for Astrophysics, School of Physics and Astronomy, University of Manchester, Oxford Road, Manchester M13 9PL, UK}
\author[0000-0002-9483-7164]{Robin Tre{\upshape{\ss}}}
\affiliation{Universit\"at Heidelberg, Zentrum f\"ur Astronomie, Albert-Ueberle-Str. 2, D-69120 Heidelberg, Germany}
\author{Monika Ziebart}
\affiliation{KOSMA, I. Physikalisches Institut, Universit\"at zu K\"oln, Z\"ulpicher Stra\ss e 77, 50937\\ K\"oln, Germany} 

\submitjournal{ApJ}
\correspondingauthor{Jorge L. Pineda}
\email{Jorge.Pineda@jpl.nasa.gov}



\begin{abstract}

  We present the first complete, velocity--resolved [C\,{\sc ii}]
  158$\mu$m image of the M51 grand-design spiral galaxy, observed with
  the upGREAT instrument on SOFIA.  [C\,{\sc ii}] is an important
  tracer of various phases of the interstellar medium (ISM), including
  ionized gas, neutral atomic, and diffuse molecular regions.  We
  combine the [C\,{\sc ii}] data with H\,{\sc i}, CO, 24$\mu$m dust
  continuum, FUV, and near--infrared K--band observations to study the
  evolution of the ISM across M51's spiral arms in both
  position--position, and position--velocity space.  Our data show
  strong velocity gradients in H\,{\sc i}, $^{12}$CO, and [C\,{\sc
      ii}] at the locations of stellar arms (traced by K--band data)
  with a clear offset in position--velocity space between upstream
  molecular gas (traced by $^{12}$CO) and downstream star formation
  (traced by [C\,{\sc ii}]). We compare the observed
  position--velocity maps across spiral arms with synthetic
  observations from numerical simulations of galaxies with both
  dynamical and quasi--stationary steady spiral arms that predict both
  tangential and radial velocities at the location of spiral arms. We
  find that our observations, based on the observed velocity gradients
  and associated offset between CO and [C\,{\sc ii}], are consistent
  with the presence of shocks in spiral arms in the inner parts of M51
  and in the arm connecting the companion galaxy, M51b, in the outer
  parts of M51.


\end{abstract}

\keywords{ISM: molecules --- ISM: structure}

\section{Introduction}
\label{sec:introduction}

The cycling of the interstellar medium (ISM) through different phases,
including the eventual formation of stars in gravitationally bound
regions, is the driving agent in the evolution of galaxies.  The
standard picture of the phases of the ISM in spiral galaxies posits
that Giant Molecular Clouds (GMCs) are assembled in the spiral arm
shocks from diffuse inter-arm H\,{\sc i} gas, and then
photo--dissociated back into the atomic phase by OB star formation
within the spiral arms \citep{Binney1998}. This picture predicts a
rapid phase change of gas across spiral arms--from the atomic to
molecular and back into atomic--synchronized by spiral arm forcing
\citep{Klessen2016}. On the other hand, CO imaging of M51 shows much
less gas-phase variation across the spiral arms \citep{Koda2009}; the
majority of the ISM gas remains molecular through the interarm
regions, surviving to the next spiral arm passage.  These observations
are consistent with numerical simulations of galactic disks, in which
a large reservoir of molecular gas is predicted to exist in the
inter--arm region of galaxies \citep{Smith2014, Duarte-Cabral2015}.

To understand galactic disks, we need to gain a full understanding of
spiral structure -- the interrelation between all the gaseous and
stellar components, and their connection to the star formation
process.  Although we can separate the different stellar populations,
measure kinematics, and study much of the ISM on a cloud--by--cloud
basis with optical, H\,{\sc i} and CO observations, we do not have the
same information on the diffuse atomic, atomic--molecular transition \citep{Wannier1991},
and warm ionized components of the interstellar medium.  These
components are traced by the [C\,{\sc ii}] 158$\mu$m line, which only now
can be observed with the velocity resolution required to understand
its link to the other phases.  [C\,{\sc ii}] traces the diffuse
ionized medium, warm and cold atomic clouds, clouds in transition from
atomic to molecular, and dense and warm photon dominated regions
(PDRs). In particular, this line is a tracer of the CO--dark H$_2$ gas
\citep{Grenier2005, Wolfire2010, Langer2010} which is likely a
precursor of the dense molecular gas that will eventually form stars.

In this paper we present the first complete, velocity resolved
[C\,{\sc ii}] 158$\mu$m map of the M51 grand-design spiral galaxy,
observed using the upgraded German REceiver for Astronomy at Terahertz
frequencies (upGREAT) instrument on the Stratospheric Observatory For
Infrared Astronomy (SOFIA).  This map was obtained as part of a Joint
Impact Proposal (program ID {\tt 04\_0116}) that also includes a
velocity unresolved [C\,{\sc ii}] map of M51 obtained with the Far
Infrared Field-Imaging Line Spectrometer (FIFI-LS) instrument on
SOFIA, and presented in \citet{Pineda2018}.  M51 is a nearby grand
design spiral at 8.6\,Mpc \citep{McQuinn2016} with a low inclination
angle of 24\degr\ \citep{Daigle2006}.  It has been extensively studied
in many tracers, including CO \citep{Aalto1999,
  Koda2009,Schinnerer2013,Miyamoto2014}, H\,{\sc i}
\citep{Walter2008}, optical light \citep{Mutchler2005}, radio
continuum \citep{Fletcher2011,Querejeta2019}, dust continuum
\citep{MentuchCooper2012}, and dense gas tracers \citep{Bigiel2016}.
These observations have allowed the spatial separation of different
ISM constituents, including H{\sc ii} regions, OB stars, and atomic
and molecular clouds. Partial maps of M51 in [C\,{\sc ii}], at low
spectral resolution, have been presented by \citet{Nikola2001} and
\citet{Parkin2013}.  Our complete [C\,{\sc ii}] map in M51 allows us
to trace the ISM phases probing atomic, PDR, and CO--dark H$_2$ gas
over the entire disk, including both arm and inter-arm regions at high
spectral resolution.
 
In spiral galaxies, spiral density waves play a fundamental role
assembling the giant molecular clouds in which star formation takes
place. There are two competing theories of the nature of spiral arms
in isolated galaxies. In the quasi--stationary spiral structure (QSSS)
hypothesis, spiral arms are thought to be rigidly rotating,
long--lived patterns that persist over several galactic rotations (see
\citealt{Bertin1996} for a review). The spiral density wave affects
the gas flow, resulting in shocks around spiral arms, triggering phase
transitions in the ISM. Alternatively, the transient spiral hypothesis
\citep{Goldreich1965,DOnghia2013,Baba2013,Dobbs2014} suggests that
each spiral arm is a transient feature generated by the
swing-amplification mechanism. Its amplitude varies on the timescale
of epicyclic motions (a fraction of galactic rotation timescale). In
this theory, the gas flows toward the potential minimum of the spiral
arm from both sides of the arm. This model is in contrast to the gas
passage from one side of the spiral arm to the other predicted in the
density wave model.  Determining the nature of spiral arms is a
fundamental aspect in the understanding of the evolution of spiral
galaxies.

 Several observational methods have been applied to distinguish
  between different theories of the nature of spiral arms in M51,
  including searches for offsets across spiral arms between stellar
  cluster with different ages
  \citep{Dobbs2010b,Shabani2018,Chandar2017} and between images of
  different ISM and star formation tracers
  \citep{Tamburro2008,Foyle2011,Louie2013,Egusa2017}. These methods,
  however, often provide contradicting results.  It has been recently
proposed that the kinematic information of gas tracers in spirals
provides an important tool for distinguishing between these two spiral
structure theories \citep{Baba2016}. Because spiral arms dynamically
affect the flow of gas, they also affect the structure of the
interstellar medium. Thus, having a complete picture of the ISM phases
with high spectral resolution observations over large areas in
galaxies is a fundamental requirement for determining the nature and
effects of spiral arms in disk galaxies. The grand--design spiral
structure in M51 is thought to be caused by the tidal interaction
between its companion galaxy M51b
\citep{Toomre1972,Pettitt2017,Tress2019}.  M51 is therefore an
excellent laboratory for studying the nature of spiral arms, and how
they affect the evolution of the ISM and star formation.
 

%
This paper is organized as follows. In
Section\,\ref{sec:observations}, we describe the [C\,{\sc ii}]
observations and the ancillary data used to study the evolution of the
ISM in M51's spirals. In Section~\ref{sec:distribution-c-sc}, we
describe techniques used to remove the rotation velocity of the galaxy
and mask the galaxy so that data can be combined to amplify the signal
produced by gas velocity variations due to the presence of spiral
density waves. In Section~\ref{sec:discussion}, we study the spatial
(2D) distribution of the different ISM traces across spiral arms
(Section~\ref{sec:aver-across-spir}) and we investigate the
distribution of ISM phases in the position--velocity space
(Section~\ref{sec:veloc-distr-across}), including a comparison between
observations and predictions from QSSS and transient spiral hypothesis
model--predicted position--velocity maps
(Section~\ref{sec:comp-with-theor}).  We summarize our results in
Section\,\ref{sec:conclusions}.


\section{Observations}
\label{sec:observations}

\begin{figure*}[t]
\centering
\includegraphics[angle=0,width=1.0\textwidth,angle=0]{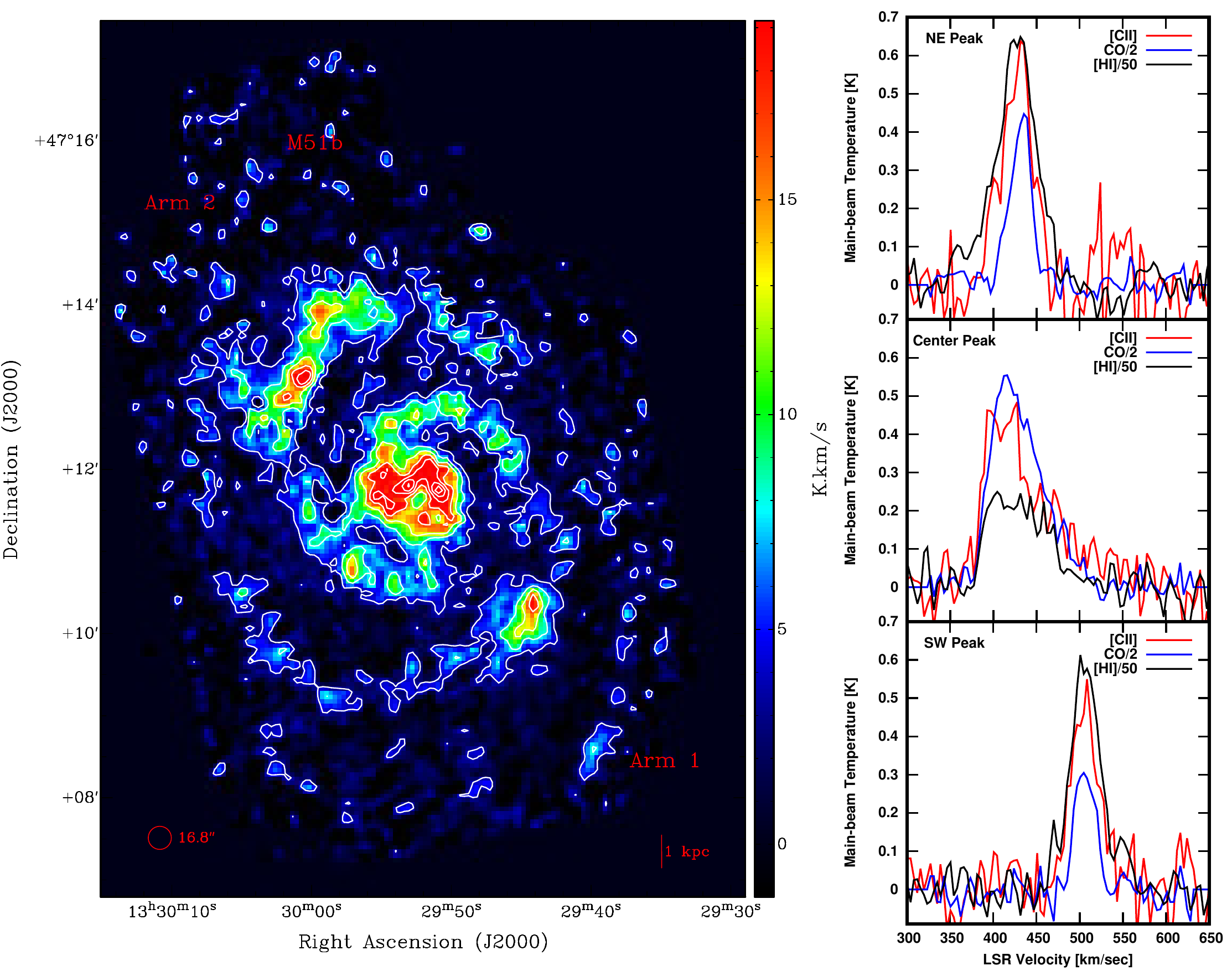}
\caption{({\it left}) [C\,{\sc ii}] integrated intensity map of M51
  observed with the upGREAT instrument on SOFIA, having an angular
  resolution of 16.8\arcsec, corresponding of 700\,pc for a distance
  to M51 of 8.6\,Mpc. The contours range from 31\% to 91\% of the peak
  integrated intensity (19.1 K km\,s$^{-1}$) in steps of 10\%.  We
  denote the location companion galaxy, M51b, in the northern part of
  the map. ({\it right}) Sample [C\,{\sc ii}], CO, and H\,{\sc i}
  spectra corresponding to the [C\,{\sc ii}] peaks at the north--east,
  central, and south--west regions in M51.  } \label{fig:m51_cii_maps}
\end{figure*}

\subsection{upGREAT Observations}
\label{sec:upgreat-observations}

We observed the [C\,{\sc ii}] $^2$P$_{3/2}-^2$P$_{1/2}$
fine--structure line at 1900.5469\,THz (rest frequency) in M51 with
the upGREAT\footnote{upGREAT is a development by the MPI f\"ur
  Radioastronomie and the KOSMA/Universit\"at zu K\"oln, in
  cooperation with the MPI f\"ur Sonnensystemforschung and the DLR
  Institut f\"ur Planetenforschung.} \citep{Risacher2016} instrument
on the Stratospheric Observatory for Far--Infrared Astronomy (SOFIA;
\citealt{Young2012}).  We covered an area of
6\arcmin$\times$12\arcmin, extending over the full extent of M51 and
its companion M51b. The upGREAT instrument uses a $2\times7$--pixel
sub--arrays in orthogonal polarization, each in an hexagonal array
around a central pixel. We employed an optimized on--the--fly (OTF)
mapping scheme in which the array is rotated by 19.1\degr\ and a
vertical and a horizontal scan are undertaken. This results in a
fully--sampled 73\arcsec$\times$73\arcsec\ square region (tile). To
cover the full extent of M51 we observed 34 such tiles.  For baseline
stability we used chopped--OTF mapping mode with two reference
positions located to the east (13:29:13.5, 47:07:32.9; J2000) and west
(13:29:31.9, 47:09:12.9; J2000) sides of M51.  The angular resolution
of the [C\,{\sc ii}] observations is 16.8\arcsec, corresponding to
700\,pc for a distance to M51 of 8.6\,Mpc.   In our analysis, we
  deprojected the [C\,{\sc ii}] cube, and all other data cubes and
  images described below, assuming the center, inclination, and
  position angle listed in Table\,1.

We processed the data using the {\tt CLASS90}\footnote{{\tt
    http://www.iram.fr/IRAMFR/GILDAS}} data analysis software. We
fitted polynomial baselines (typically of order 3), smoothed the data
in velocity, and resampled the data into a regular spatial grid. We
also apply a set of main--beam efficiencies that are appropriate to
each pixel of the upGREAT array to transform the data from antenna
temperature to main--beam temperature scale. The average main--beam
efficiency is 0.68 \citep{Risacher2016}. The typical rms noise of the
resulting data is 0.06\,K in a 3.8\,km\,s$^{-1}$ channel width. We
compared the integrated intensities of our upGREAT [C\,{\sc ii}] map
with those observed in the inner parts of M51 with Herschel/PACS
\citep{Parkin2013}, smoothed to a 16.8\arcsec\ angular resolution, and
with the FIFI--LS [C\,{\sc ii}] map presented in
\citet{Pineda2018}. We find good agreement in the integrated
intensities between these three maps with differences within the
uncertainties of the observations.  The integrated intensity map of
M51 observed with the upGREAT instrument is shown in the left panel of
Figure~\ref{fig:m51_cii_maps}. The [C\,{\sc ii}] distribution in M51
is characterized by bright emission in its central region, with the
[C\,{\sc ii}] peak being about 22.5\arcsec\ or 900\,pc from M51's
center, and in two regions at about 5.3\,kpc from the galaxy's center.
In the right panel of Figure~\ref{fig:m51_cii_maps}, we show the
velocity--resolved [C\,{\sc ii}], $^{12}$CO, and H\,{\sc i} spectra
corresponding to the [C\,{\sc ii}] peaks at the north--east, central,
and south--west regions. We see that the line [C\,{\sc ii}] widths are
typically broader than those of CO, but somewhat narrower compared
with those of H\,{\sc i}.  The companion galaxy, M51b, which is
located at the northern portion of the map, is not detected in our
upGREAT map, but it is detected in our FIFI--LS map.  A comparison
between H$\alpha$ and mid-- and far--infrared tracers and [C\,{\sc
    ii}] is discussed in \citet{Pineda2018}.

\subsection{H\,{\sc i} and CO Observations}

In our analysis we employed the H\,{\sc i} spectral cube obtained in
M51 using the VLA as part of the THINGS project
\citep{Walter2008}. The data has been produced with a robust weighting
scheme and has an angular resolution of $\sim$6\arcsec. We smoothed
the data to 16.8\arcsec\ and regridded the H\,{\sc i} data to that of
the [C\,{\sc ii}] map for comparison. Note that these interferometric
observations lack short spacing data. However, these observations are
sensitive to scales up to $\sim$15\arcmin, greater than the full
extent of M51. Additionally, the total H\,{\sc i} flux of M51 observed
with the VLA is consistent with previous, lower resolution,
single--dish observations \citep[see Table 5 in ][]{Walter2008}. The
rms noise of the H\,{\sc i} data is 0.3\,K in a 3.8\,km\,s$^{-1}$
channel width.

We used the $^{12}$CO $J=1\to0$ map observed with the CARMA
interferometer, with data for a short spacing correction obtained with
the  Nobeyama Radio Observatory 45 telescope \citep{Koda2009}. We
smoothed and regrided the $^{12}$CO map, from its original resolution
of 4\arcsec, to match the 16.8\arcsec\ resolution of the [C\,{\sc
    ii}] data cube. The resulting $^{12}$CO maps has a rms noise of
0.07\,K in a 3.8\,km\,s$^{-1}$ wide channel.
We also employed the $^{12}$CO and $^{13}$CO $J=1\to0$ cubes observed
with the IRAM 30\,m telescope presented by \citet{Pety2013}, as they
better sample regions in the outer portions of M51.  The angular
resolution of the $^{12}$CO and $^{13}$CO cubes is 22.5\arcsec\ and
the rms noise is 0.016\,K in a 5\,km\,s$^{-1}$ wide channel.

\subsection{24$\mu$m continuum, FUV, and K--band near-IR Maps}

We also compared our [C\,{\sc ii}] dataset with 24$\mu$m, FUV maps,
and K--band optical maps, tracing warm dust, evolved unobscured star
formation, and stellar mass, respectively.  The 24$\mu$m continuum map
was observed as part of the {\it Spitzer}/SAGE survey
\citep{Kennicutt2003} and has a native angular resolution of
6\arcsec. The FUV image was obtained by the GALEX satellite and
presented by \citet{GildePaz2007} with a native resolution of
4.2\arcsec. The K--band near--infrared data was observed by 2MASS and
presented by \citet{Jarrett2003} with a native resolution of
2.5\arcsec. All maps were smoothed with a Gaussian kernel and
regridded to match the 16.8\arcsec\ angular resolution and positions of
our [C\,{\sc ii}] observations.

\section{Gridding,  Masking, and Derotation}
\label{sec:distribution-c-sc}

\subsection{Spatial Gridding}
\label{sec:mask-definition}

\begin{deluxetable}{lc} 
\tabletypesize{\scriptsize}
\centering \tablecolumns{2} \small
\tablecaption{Adopted Parameters of M51} \tablenum{1}
\tablehead{\colhead{Parameter} \hspace{2cm} & \colhead{Value}}
\startdata
R.A. (J2000) & 13:29:52.771 \\ Decl. (J2000) & +47:11:42.62
\\ P.A.$^{1}$ & 169.0$\pm$4.2\degr \\ Incl.$^{2}$ & 24$\pm$3\degr
\\ Distance & 8.6\,Mpc \enddata \tablenotetext{1}{Position angle taken
  from \citet{Shetty2007}.} \tablenotetext{2}{Inclination taken from
  \citet{Daigle2006}.}
\end{deluxetable}

In the following sections we discuss the distribution of different ISM
and star formation tracers across the spiral arms of M51 in both
spatial and spatial--velocity space. In order to allow the comparison
with continuum tracers (FUV, K--band NIR, etc), we first studied the
CO and [C\,{\sc ii}] integrated intensities in our analysis. We use
the azimuthal distribution of these intensities to investigate
possible shifts in their peak intensity that would result from the
presence of spiral density waves. Such azimuthal intensity
distributions are often averaged in a set of annuli with a given
radial width with the aim of improving the signal--to--noise ratio of
the observations.  However, within such annuli, the spiral arm
structure varies rapidly with radius, and averaging in the radial
direction results in confusion between arm and inter--arm regions. To
facilitate the separation between arm and inter--arm regions,
following \citet{Koda2012}, we defined a set of spiral arm segments
that follow the spiral structure of M51.  Such segments ensure that
the averaging is done in arms and inter--arm regions separately.  The
spiral structure of a galaxy can be characterized by logarithmic
spirals given as,
\begin{equation}\label{eq:2}
\theta=\frac{-1}{\tan(i_{\rm pitch})}\ln(r)+\theta_0,
\end{equation}
where $\theta$ is the azimuthal angle, $r$ the radius, and $i_{\rm
  pitch}$ the pitch angle, and $\theta_0$ is the phase angle. The
origin of azimuthal angle starts at the west part of the map and
increases clockwise.  The azimuthal angle is the counterclockwise
angle from the positive x-axis (west) in the map.  The entire extent
of M51 cannot be described by a single pitch angle, as changes in this
angle are apparent in the two [C\,{\sc ii}] bright regions about
127\arcsec\ (5.3\,kpc) from the center, and at the outer spiral
arms. We therefore define four regions that are characterized by
different values of $i_{\rm pitch}$: M51's inner galaxy, middle
north, middle south, and Outer galaxy. The
radial limits and pitch angles of these regions are listed in Table
2.  For the inner galaxy, we adopted the pitch angle used by
  \citet{Koda2012}, but for the middle and outer masks we adopted
  pitch angles that approximately match the observed spiral
  pattern. We tested whether small variations ($\pm$10\%) in the
  assumed pitch angles affect the resulting spatial and
  spatial--velocity distributions studied below and found that they
  are not significantly affected.

\begin{figure*}[t]
\centering
\includegraphics[angle=0,width=0.75\textwidth,angle=0]{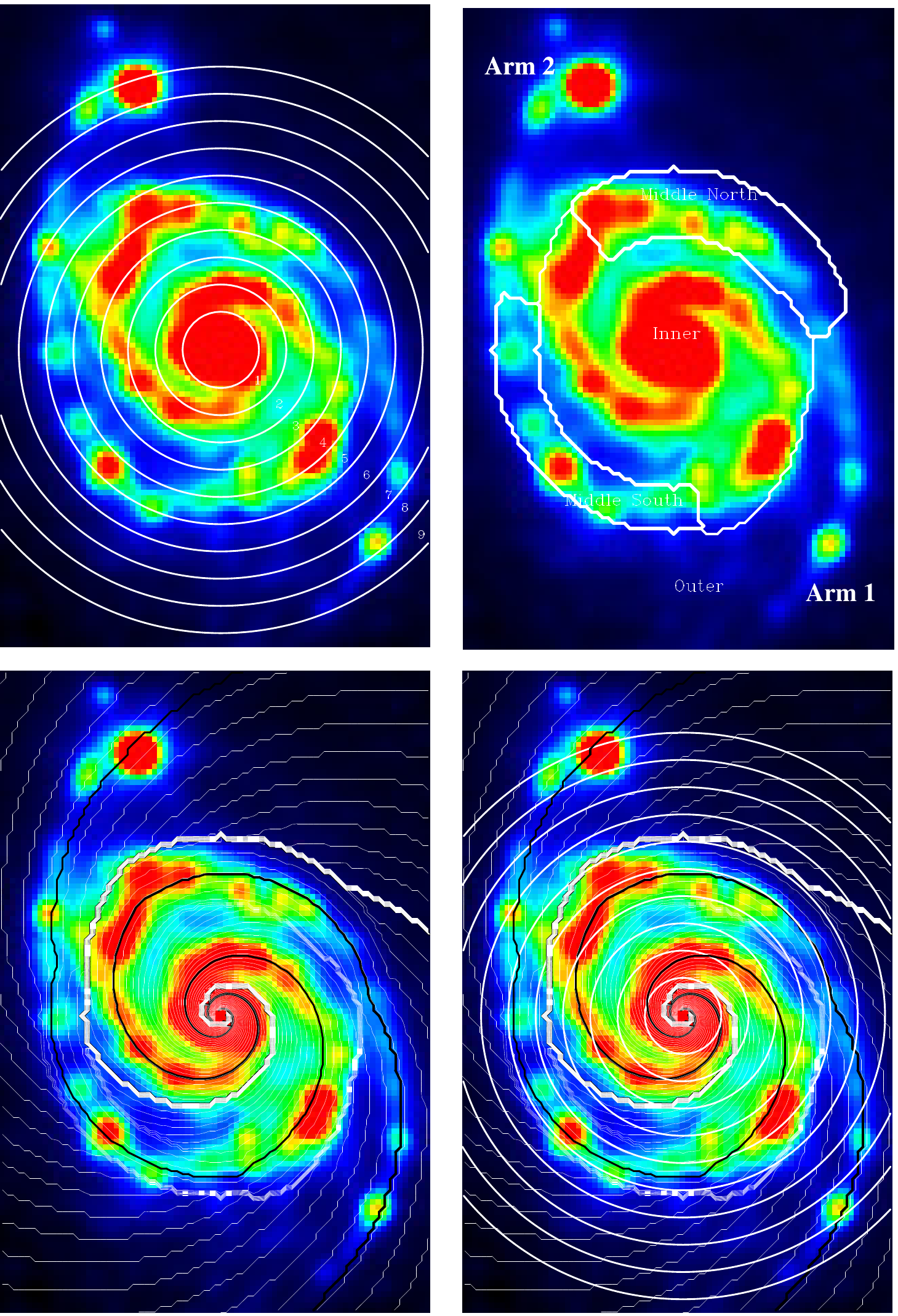}
\caption{({\it top left}) Azimuthal rings displayed over the 24$\mu$m
  continuum emission image of M51.  The rings extend from 1.5\,kpc to
  9.1\,kpc, in steps of $\sim$1\,kpc. ({\it top right}) Definition of
  the regions that are characterized by different values of the pitch
  angle (see Table 2) that are used to create the spiral pattern
  grid. ({\it bottom left}) Logarithmic spiral pattern definition
  overlaid on the 24$\mu$m continuum emission image of M51 (see
  Section~\ref{sec:distribution-c-sc}). The black lines show the
  spiral pattern at the location of the spiral arms and the thick
  white line counterclockwise of Arm 2 represents the origin,
  $\theta_0=0$ rad, of the phase angle distribution. ({\it bottom
    right}) The combined azimuthal and spiral mask that defines the
  segments from which we derive averaged intensities and the stacked
  spectra used in our analysis.}\label{fig:definition}
\end{figure*}

We construct a grid of spiral segments over M51 that are parameterized
by the phase angle $\theta_0$ in Equation~(\ref{eq:2}). We scale the
values of $\theta_0$ in each mask to range between 0 and 2$\pi$, and
we rotate the origin of the phase angle distribution so that it
follows the spiral pattern. This rotation results in the spiral arms
being at about the same phase angle at any radius.
In Figure~\ref{fig:definition} we illustrate our definition of the
spiral arm segments used in our analysis over the 24\,$\mu$m dust
continuum image of M51. In the upper left panel, we show the radial
mask, which corresponds to rings of $\sim1$\,kpc width extending from
1.5\,kpc and 9.1\,kpc. In the upper right panel, we show the
definition of the Inner, Middle North, Middle South, and outer regions which
are characterized by different pitch angles (Table\,2). In the lower
left panel, we show the spiral grid, which is defined by 32 segments
with a step in $\theta_0$ of 0.098\,radians. We highlight in black the
spiral segments at the location of the spirals in the 24\,$\mu$m
image. The thick white lines are artifacts of the contouring that
becomes thicker when there are discontinuities in the phase angle
distribution. Such discontinuities are either produced at the origin
of the phase angle distribution, $\theta_0=0$ rad, or at the middle
masks where spiral arms are close to each other due to the change in
the pitch angle. The thicker white line that denotes the origin of the
phase angle distribution at each radius follows Arm 2 in the
counterclockwise side.  Finally, in the lower left panel we show the
combined radial and spiral masks in M51.

\begin{figure}[t]
\centering
\includegraphics[angle=0,width=0.45\textwidth,angle=0]{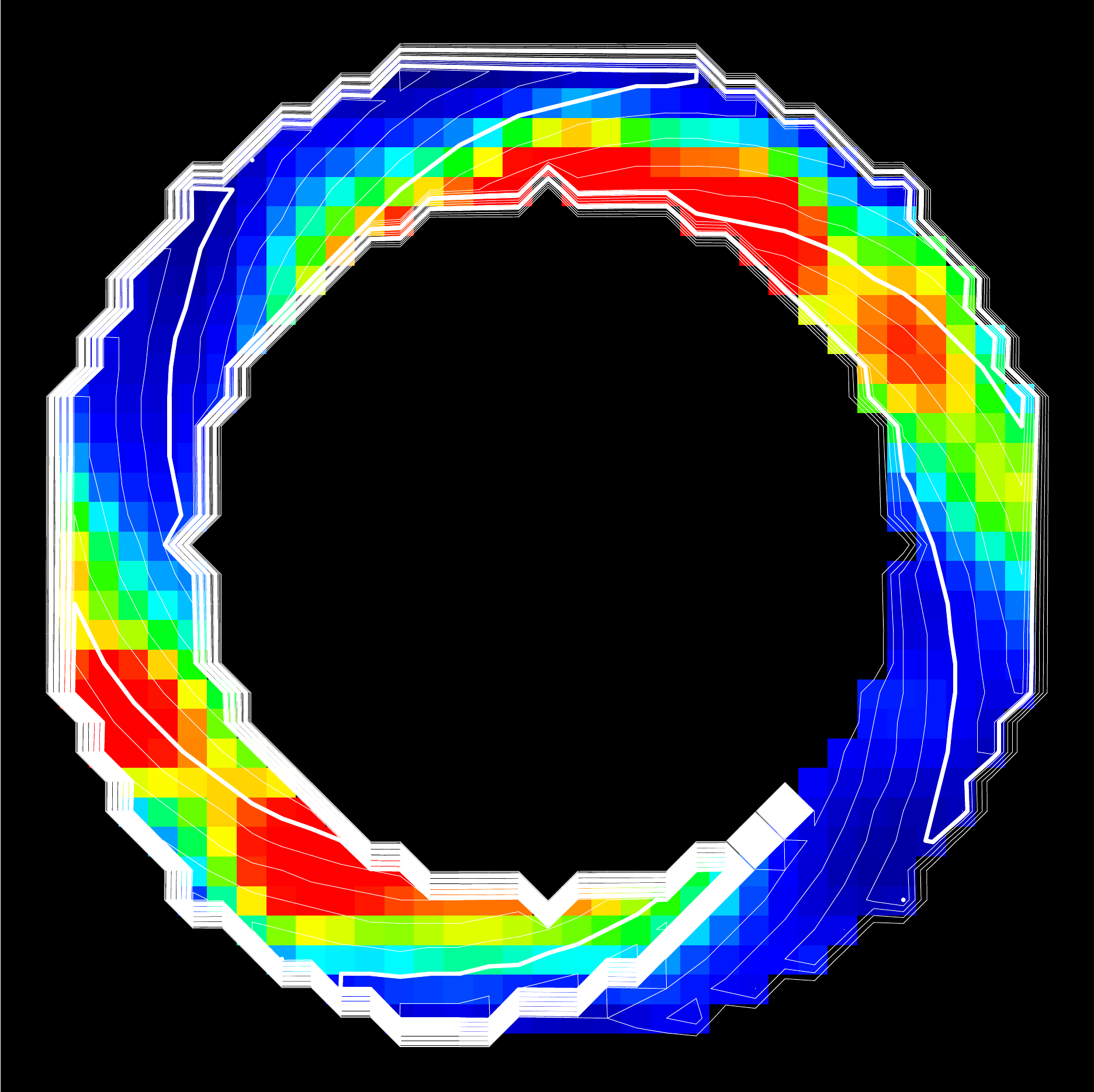}
\caption{Definition of spiral segments for a ring between 2.6\,kpc and
  3.6\,kpc from M51's center, overlaid over the 24\,$\mu$m dust
  continuum image on M51.  The thick white line at the bottom
  represents the origin, $\theta_0=0$ rad, of the phase angle
  distribution which increases anti-clockwise.}\label{fig:spiral_zoom}
\end{figure}

  We show a detailed view of the spiral segments in
  Figure~\ref{fig:spiral_zoom}. The spiral segments shown in the
  figure correspond to Ring 2, which extends between 2.6\,kpc and
  3.6\,kpc from M51's center. The integrated intensity and
  position--velocity distribution discussed in
  Section~\ref{sec:veloc-distr-across} are the result of the data
  averaged within each spiral segments in the figure. The thick white
  line at the bottom represents the origin, $\theta_0=0$ rad, of the
  phase angle distribution which increases counterclockwise. This
  method has the advantage that it does not depend on any specific
  definition of a spiral arm in the image data and ensures that
  emission in arm and inter-arm regions are not averaged together.  



\begin{figure}[h]
\centering
\includegraphics[angle=0,width=0.5\textwidth,angle=0]{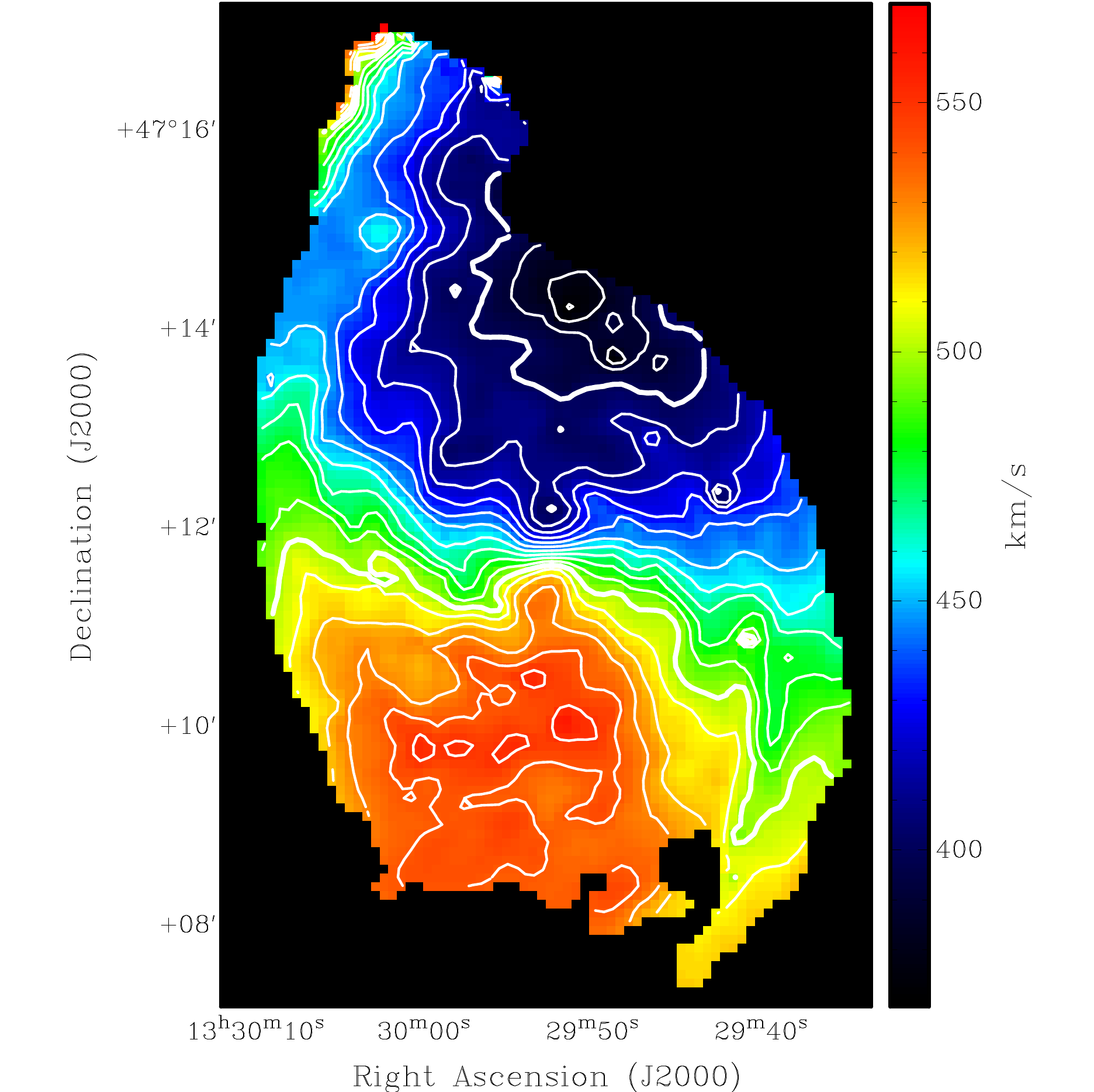}
\caption{Mass--weighted velocity centroid distribution in the disk of
  M51 derived from H\,{\sc i} and $^{12}$CO maps. Contours range from
  360 to 570 km\,s$^{-1}$ in steps of 10
  km\,s$^{-1}$.}\label{fig:vel_field}
\end{figure}

\begin{deluxetable}{lccccc} 
\tabletypesize{\scriptsize} \centering \tablecolumns{6} \small
\tablewidth{0pt}
\tablecaption{Definition of Regions in  M51}
\tablenum{2}
\tablehead{\colhead{Region} & \colhead{$R^{1}_{\rm in}$}  & \colhead{$R^{1}_{\rm out}$}  & \colhead{$\Phi^{2}_{\rm in}$} & \colhead{$\Phi^{2}_{\rm out}$} & \colhead{$i^{3}_{\rm pitch}$} }   
\startdata
Inner$^4$    & 0\arcsec\ & 160\arcsec & 0 & 2$\pi$  & 18.5\degr \\
Middle North$^5$  & 106\arcsec\  & 180\arcsec & 0 & 0.7$\pi$ & 2\degr \\
Middle South$^5$  & 132\arcsec\  & 191\arcsec & 0.95$\pi$  & 1.55$\pi$ & 2\degr \\
Outer$^4$    & 160\arcsec\ & 300\arcsec & 0 & 2$\pi$ & 28\degr \\
\enddata
\tablenotetext{1}{Radius from the center of M51.\\}
\tablenotetext{2}{Azimuthal Angle.}
\tablenotetext{3}{Pitch Angle.}
\tablenotetext{4}{Excluding regions contained in Middle North and South masks.}
\tablenotetext{5}{Small adjustments were made to the radial range to optimize the correspondence between the spiral binning and the intensity distribution in M51.}
\end{deluxetable}

\begin{figure*}[t]
\includegraphics[width=0.95\textwidth,angle=0]{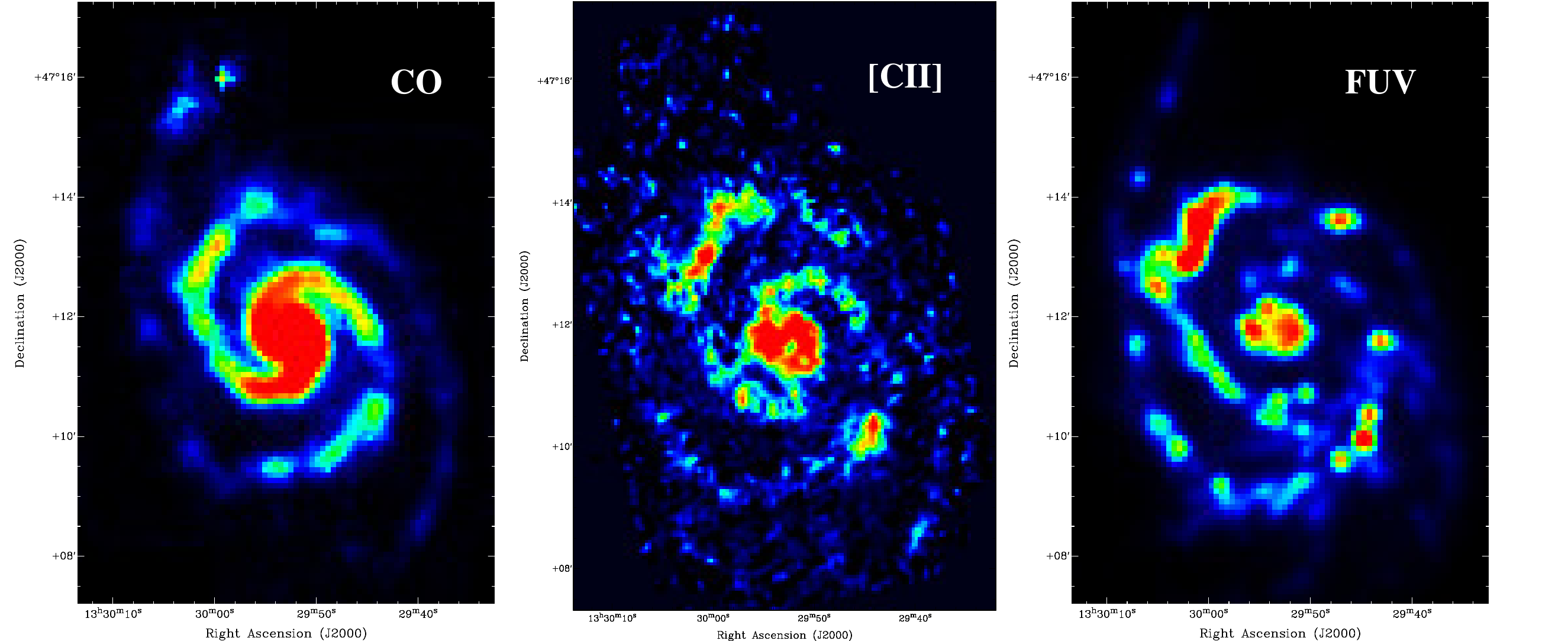}
\centering
\caption{$^{12}$CO ({\it left}), [C\,{\sc ii}] ({\it center}), and FUV
  ({\it right}) images of M51. The $^{12}$CO, [C\,{\sc ii}], and FUV
  maps trace cold gas, warm and dense PDRs, and evolved unobscured
  star forming regions, respectively.  }\label{fig:cii_co_fuv}
\end{figure*}

\subsection{Velocity Derotation}
\label{sec:velocity-derotation}


Spiral density waves not only can result in different spatial
distributions of CO, H\,{\sc i}, [C\,{\sc ii}], and other tracers, but
they also imprint a particular velocity distribution on the gas
\citep{Roberts1987,Baba2016}. These velocity features, if related to
the compression of gas caused by shocks, are likely associated with
the transition between ISM phases, and therefore variations in the
position--velocity distribution of CO, H\,{\sc i}, and [C\,{\sc ii}]
are expected.

The observed velocity of spectral lines in the galaxy is a combination
of M51's systemic velocity, the rotational velocity, and peculiar
(related to the spiral perturbation) velocities \citep[see
  e.g.][]{Shetty2007}.  With the aim to study how the
position--velocity distribution of H\,{\sc i}, CO, and [C\,{\sc ii}]
intensities is affected by spiral density waves, we need to isolate
velocity variations due to spiral density waves from those due to the
systemic and rotation velocity of the galaxy. To do so, we define a
reference velocity frame that co--rotates with the spiral density
perturbation.  Assuming that the gas mass peaks in an area associated
with the location of the spiral density perturbation, we define the
velocity at which most of the mass rotates to be the reference
velocity frame. We start   
by estimating the total hydrogen surface density per velocity bin at
each voxel in a position--velocity data cube of M51.  We generate a
total hydrogen surface density per unit velocity, $\sigma_{\rm
  H}=\sigma_{\rm H^0}+\sigma_{\rm H_2}$, cube by combining the H\,{\sc
  i} and $^{12}$CO $J=1\to0$ data cubes.  The neutral atomic, H$^0$,
and molecular, H$_2$, hydrogen surface densities per velocity bin are
given by,
\begin{equation}
\sigma_{\rm H^0}=\Sigma_{\rm H^0}/ \Delta v, 
\end{equation}
and 
\begin{equation}
\sigma_{\rm H_2}=\Sigma_{\rm H_2}/ \Delta v, 
\end{equation}
respectively, where $\Sigma_{\rm H^0}$ and $\Sigma_{\rm H_2}$ are the
H$^0$ and H$_2$, surface densities, in units of M$_\odot$\,pc$^{-2}$,
which are in turn given by,
\begin{equation}
\Sigma_{\rm H^0}=0.002 I_{\rm HI}=0.002 T_{\rm HI} \Delta v, 
\end{equation}
and
\begin{equation}
\Sigma_{\rm H_2}=4.37  I_{\rm CO}=4.37 T_{\rm CO} \Delta v,
\end{equation}
where $I_{\rm HI}$ and $I_{\rm CO}$, are the H\,{\sc i} and CO
integrated intensities in units of K\,km\,s$^{-1}$, and $T_{\rm HI}$
and $T_{\rm CO}$ intensities at a given velocity channel in units of
K.  
%
 Note that a $\cos(i)$ factor, where $i$ is the galaxy's inclination
 (Table I), needed to correct the surface brightness is applied during
 the deprojection step.  The surface density calculation includes a
 factor of 1.36 to account for heavy elements and the H$_2$ surface
 density calculation assumes a $X_{\rm CO}=2\times10^{20}$
 cm$^{-2}$(K\,km\,s$^{-1}$)$^{-1}$ \citep{Bolatto2013}.  We calculate
 the mass--weighted velocity centroid in each spatial pixel of the
 data cube using,
\begin{equation}
  V_{\rm cen}=\frac{\sum_i^N \sigma^i_{\rm H} V^i_{\rm
      H}}{\sum_i^N \sigma^i_{\rm H}},
\label{eq:1}
\end{equation}
over N channels in each spectra.   The uncertainties in the
  resulting velocity centroids are mostly related to rms noise in the
  spectra and the presence of noise peaks across the velocity band. We
  minimize these effects in the derived mass--weighted velocity
  centroid map by using a diluted mask technique, in which the data
  cube is smoothed to a lower resolution, so that noise of the spectra
  is reduced by a factor of $\sim$3. We then create a
  position--position--velocity mask in which intensities are larger
  than 4 times the rms noise of the smoothed cube, and applied it to
  the unsmoothed data cube. The resulting mass--weighted velocity
  centroid map is shown in Figure~\ref{fig:vel_field}.

%
%

We use the total hydrogen mass--weighted velocity centroid map to shift
the velocity axis of the H\,{\sc i}, CO, and [C\,{\sc ii}] spectra so
that the velocity corresponding to $V_{\rm cen}$ becomes zero.  This
results in data cubes that peak around $V=0$\,km\,s$^{-1}$, but show
deviations of about $\pm$30\,km\,s$^{-1}$.  Because these velocity
deviations are relative to the velocity of the total hydrogen
intensity peak, in a given spectra, the relative differences between
the [C\,{\sc ii}], H\,{\sc i}, and $^{12}$CO line emission are not
sensitive to the exact value of the systemic and rotational
components.  The de--rotated data cubes will be used in
Section\,\ref{sec:veloc-distr-across} to study the distribution of
H\,{\sc i}, CO, and [C\,{\sc ii}] across spiral arms.

\begin{figure*}[t]
\centering
\includegraphics[angle=0,width=0.85\textwidth,angle=0]{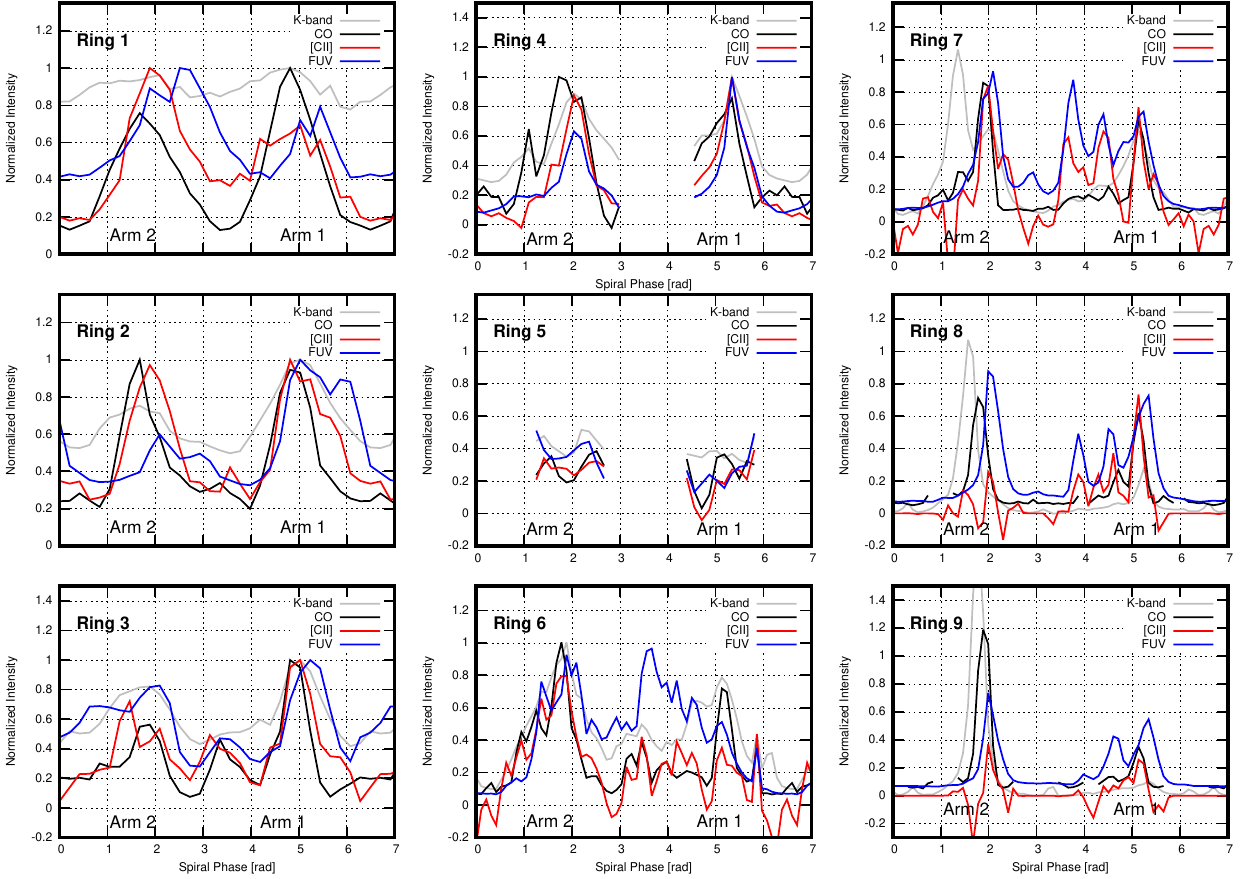}
\caption{[C\,{\sc ii}], $^{12}$CO, FUV, and K--band NIR infrared
  intensities as a function of spiral phase. Each panel represent a
  ring with $\sim$1\,kpc width extending from 1.5\,kpc to 9.1\,kpc (see
  Figure~\ref{fig:definition}).  }\label{fig:radial_avg}
\end{figure*}

\begin{figure*}[t]
\includegraphics[width=0.8\textwidth,angle=0]{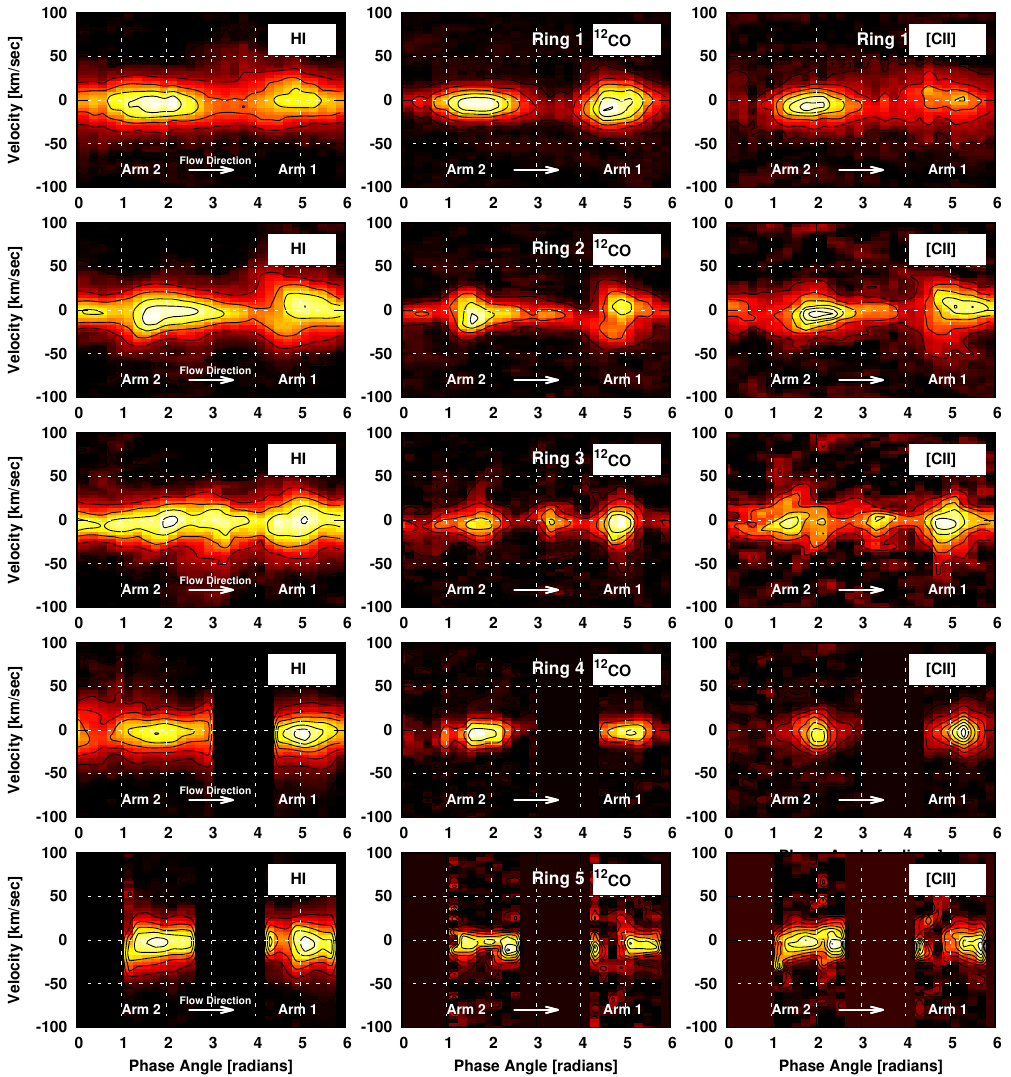}
\centering
\caption{H\,{\sc i}, $^{12}$CO, [C\,{\sc ii}] position velocity maps
  as a function of phase angle. Each panel corresponds to a
  $\sim$1\,kpc radial segment extending from 1.5\,kpc to
  6.9\,kpc. }\label{fig:spiral_pv_maps}
\end{figure*}

\begin{figure*}[t]
\includegraphics[width=0.8\textwidth,angle=0]{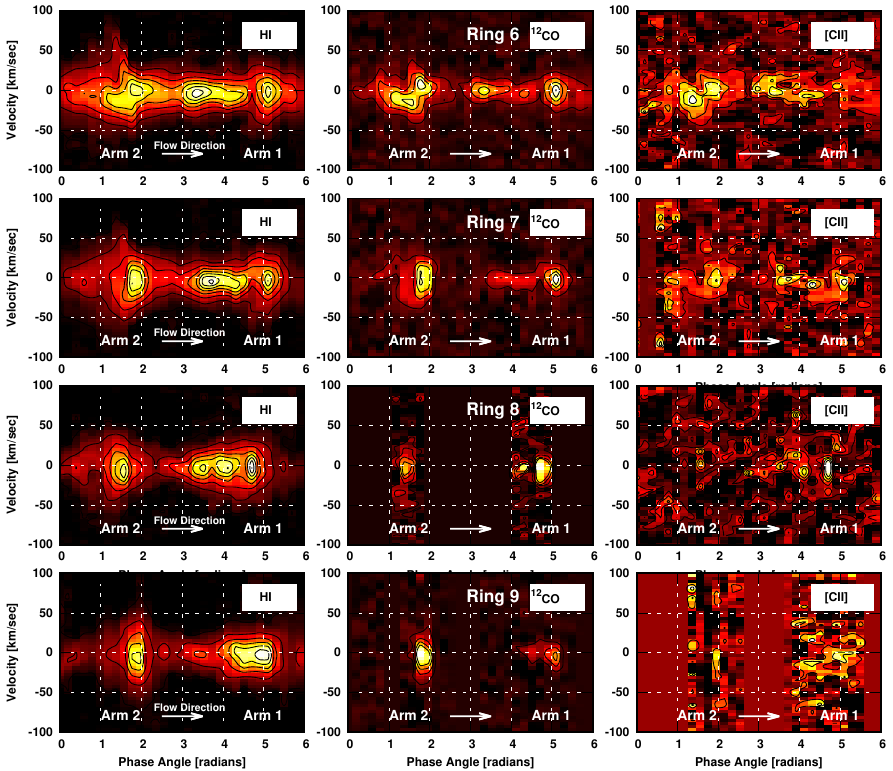}
\centering
\caption{H\,{\sc i}, $^{12}$CO, [C\,{\sc ii}] position velocity maps
  as a function of phase angle. Each panel corresponds to a
  $\sim$1\,kpc radial segment extending from 6.9\,kpc to
  11.2\,kpc. }\label{fig:spiral_pvmaps_v2}
\end{figure*}

\section{Discussion}
\label{sec:discussion}
\subsection{Integrated Intensity Distribution across Spiral Arms}
\label{sec:aver-across-spir}

The integrated intensity distribution of different tracers as a
function of azimuthal angle for a set of different annuli in galaxies
has been used to study the evolution of the ISM in spiral arms and to
test theories of spiral structure
\citep{Tamburro2008,Foyle2011,Egusa2017}.  The quasi--stationary
spiral arm structure (QSSS) theory predicts that, inside the
co-rotation radius\footnote{Defined as the location at which the
relative velocity between the gas and stars and the stellar pattern is
zero.}, the location of gaseous spiral arms move from upstream to
  downstream of stellar arms with an offset that decreases as the
  radius increases. Beyond the co-rotation radius the gaseous arms
  move from upstream to downstream with the offset increasing with
  radius \citep{Gittins2004}.  \citet{Egusa2017} find that in the
  inner M51, one arm (Arm 2; see Figure~\ref{fig:m51_cii_maps}) shows
  offsets between CO and H$\alpha$ emission that are radially ordered
  between the stellar and gas arms, while the other (Arm 1) does
  not. Additionally, offsets are not clearly seen in the outer regions
  of M51. They conclude that the nature of two inner spiral arms in
  M51 is different than of the outer arms, due to the interaction with
  the companion galaxy, M51b.  \citet{Foyle2011} studied spatial
  offsets between different ISM and star formation tracers in a sample
  of 12 galaxies (including M51) using the cross--correlation method
  \citep{Tamburro2008,Dobbs2010}.  They find no systematic offset
  variation with radius in any galaxy in their sample and conclude
  that their observations are inconsistent with the QSSS theory.  With
  the aim at resolving the discrepancies between the studies mentioned
  above, \citet{Louie2013} studied different methods for determining
  offsets between gas spiral arm and star formation tracers.  They
  argue that offsets are highly dependent on which gas tracer is used
  (e.g. CO or H\,{\sc i}) and that different methods for measuring the
  location of spiral arms (Gaussian fitting, cross--correlation) can
  give discrepant results.  They find mostly positive offsets between
  CO and H$\alpha$, suggesting that gas flow through spiral arms
  (i.e., density wave), although the spiral pattern may not
  necessarily be stationary for a timescale much longer than the
  arm-crossing timescale.

%
%

[C\,{\sc ii}] provides an unobscured view of the far--ultraviolet
illuminated regions associated with massive star formation
\citep{Pineda2018}. These observations can be combined with those of
low--$J$ transitions of CO, which traces cold and moderately dense
molecular gas, and FUV, describing evolved regions in which star
formation disrupted their progenitor molecular gas, to study the
evolution of the interstellar medium across spiral arms.

In Figure~\ref{fig:cii_co_fuv}, we show our [C\,{\sc ii}] map,
together with those of $^{12}$CO and the far--ultraviolet intensity.
We see a morphological evolution between $^{12}$CO, [C\,{\sc ii}], and
FUV, not only represented as apparent offsets along the flow direction
(counterclockwise), but also in terms of small scale structure, with
CO being more ordered, while [C\,{\sc ii}] and FUV are more
non-uniformly distributed. This difference is possibly a result of
[C\,{\sc ii}] and FUV tracing more energetic environments compared
with $^{12}$CO.

We present a more detailed view of the distribution of different
tracers across spiral arms in M51 in Figure~\ref{fig:radial_avg}.  We
show the normalized integrated intensity of $^{12}$CO, [C\,{\sc ii}],
and FUV as a function of the phase angle for a set of rings (Rings 1
to 9; see Figure \ref{fig:definition}) extending from 1.5\,kpc to
9.1\,kpc, in steps of $\sim$1\,kpc. We also include near--infrared K
band observations which trace the stellar mass, and thus the location
of bottom of the gravitational potential \citep{MentuchCooper2012}.  In
the inner galaxy (Rings 1--5), we use data at 16.8\arcsec\ resolution,
with the $^{12}$CO map presented by \citet{Koda2009}, while for the
outer galaxy (Rings 6--9) we take data smoothed to 23\arcsec,
corresponding to the angular resolution of the $^{12}$CO map presented
by \cite{Pety2013}. This data cube has better coverage and
signal--to--noise ratio in the outer regions of M51.  We mask out from
these figures the galactic center and companion galaxy, M51b, using
the mask presented by \cite{Pineda2018}.  Because of the sudden
reduction of the spiral arms pitch angle in the region between
4.75\,kpc and 6.9\,kpc (Rings 4 and 5), the inter-arm area in this
region is greatly reduced and our spiral grid corresponding to these
inter-arm regions is truncated. This results in the gaps seen in
Figure~\ref{fig:radial_avg} and Figure~\ref{fig:spiral_pv_maps} for
these regions.  We see peaks that correspond to the two spiral arms
with the direction of the flow being left to right.
%
%
%
Following \cite{Egusa2017}, we define the arm on the left, connecting
to the companion galaxy M51b, as Arm 2 and that on the right, pointing
away from the companion, as Arm 1 (see Figure~\ref{fig:m51_cii_maps}).

In the inner M51 (Rings 1 to 3 in Figure~\ref{fig:radial_avg}), there
is a clear offset between the peaks of $^{12}$CO, [C\,{\sc ii}], and
FUV for Arm 2, but that is not as pronounced in Arm 1, in particular
between $^{12}$CO and [C\,{\sc ii}].  The [C\,{\sc ii}] and FUV
  profiles, however, show more extended emission in the direction of
  the flow compared with CO for both arms. The missing offset between
  in the intensity peaks in Arm 1 is consistent with that seen by
  \citet{Egusa2017} but, as we will see in
  Section~\ref{sec:veloc-distr-across}, it is likely the result of
  using integrated intensities lacking spectral information.  The
stellar arms traced by NIR K--band emission are typically associated
with the gas peaks, but there is no significant offset between them.
In the middle region of M51 (Rings\,4 and 5), where the pitch angle of
spiral arms is significantly reduced, we can still see an offset
between $^{12}$CO, [C\,{\sc ii}], and FUV.  In the outer galaxy,
(Rings 6 to 9) the offsets appear to be smaller. Note that the peak in
NIR K--band in Ring\,9 for Arm\,2 is associated with emission from the
companion galaxy that is outside M51b' mask.

In Rings 6--8 (see also Figure~\ref{fig:sample_pv_2}), we see emission
peaks in the inter--arm regions that are prominent in [C\,{\sc ii}]
and FUV, but are relatively faint in CO emission. These peaks
correspond to arm--like structures in the inter--arm regions in the
southern part of M51. The large [C\,{\sc ii}]/$^{12}$CO ratio (in
units of K\,km\,s$^{-1}$) is suggestive of the presence of CO--dark
H$_2$ gas in the region \citep{Pineda2013}. A quantitative study of
the inter-arm [C\,{\sc ii}] in M51 will be presented in a subsequent
publication in this series.

Due to the 700\,pc spatial resolution of our observations,
insufficient for resolving spiral arms spatially, we are unable to
confirm whether there is a systematic variation in the offsets between
NIR K--band, $^{12}$CO, [C\,{\sc ii}], and FUV as a function of
radius. We can, however, use the high spectral resolution of our
observations to study the nature of spiral arms in M51, as discussed
in the following Section.



\subsection{Velocity Distribution across Spiral Arms}
\label{sec:veloc-distr-across}

    The two proposed theories of spiral arm formation discussed above
    make different predictions about the velocity field of the gas
    encountering a spiral arm.  In the QSSS hypothesis the gas
    velocity changes suddenly as it approaches the spiral arm, while
    in the dynamic spiral hypothesis the gas flows fall into the
    spiral arm from both sides with no, or little, shock.  Because
    spiral structure dynamically affects the flow of gas, the ISM in
    spiral arms transitions from diffuse atomic to dense molecular
    gas, followed by star formation. Thus, as discussed above, the
    shocks suggested by the QSSS theory result in a systematic offset
    between gas and star formation tracers that is not expected in the
    dynamic spiral theory. Shocks are also predicted in certain spiral
    arm regions of tidally interacting systems like M51
    \citep{Oh2008,Dobbs2010,Oh2015,Pettitt2017}
 
      \citet[][see also \citealt{Roberts1987}]{Baba2016}, proposed
      that the kinematic information of gas tracers in spirals
      provides an important tool for distinguishing between these two
      spiral structure theories. They used hydrodynamic simulations to
      study the tangential and radial velocities of the gas in two
      simulated galaxies with quasi-stationary and dynamical
      spirals. They find that the tangential and radial velocities in
      the quasi-stationary spiral model show a relatively large
      velocity gradient with a well defined pattern that repeats as
      the gas flow encounters a spiral arm
      (Figure~\ref{fig:velocity_theory}; see also Figure 4 in
      \citealt{Baba2016}). In contrast the galaxy with dynamic spirals
      shows a less defined pattern with more moderate velocity
      variations at the location of spiral arms.  In the following, we
      will use the position--velocity structure of the CO and [C\,{\sc
          ii}] gas across the spiral arms to study the nature of M51's
      spirals.

 In Figure~\ref{fig:spiral_pv_maps} and \ref{fig:spiral_pvmaps_v2}, we
 show the H\,{\sc i}, $^{12}$CO, and [C\,{\sc ii}] position--velocity
 maps for $\sim$1\,kpc wide rings in the inner and outer M51,
 respectively. Each spectrum in the position--velocity maps
 corresponds to the average within a spiral segment defined in
 Section~\ref{sec:mask-definition} (see also
 Figure~\ref{fig:spiral_zoom}).  The two intensity peaks correspond to
 M51's spiral arm locations and the gas flows from left to right. As
 discussed in Section~\ref{sec:aver-across-spir}, in
 Figure~\ref{fig:spiral_pv_maps} we use data at
 16.8\arcsec\ resolution, while for Figure~\ref{fig:spiral_pvmaps_v2},
 we use data smoothed to 23\arcsec.  In two locations (Rings 2 and 6)
 we see a velocity gradient in all tracers, which is more pronounced
 in $^{12}$CO. In Ring 2 we see an antisymmetric velocity pattern for
 Arm 1 and 2.  Offsets in the position--velocity space between
 $^{12}$CO and [C\,{\sc ii}] are noticeable. The H\,{\sc i} emission
 appears to be less structured in the inner galaxy compared with the
 outer galaxy. In general, H\,{\sc i} extends across the spiral arms
 coinciding with both the $^{12}$CO and [C\,{\sc ii}] emission. It has
 been suggested that H\,{\sc i} emission in spiral galaxies traces
 both diffuse atomic clouds and gas photodissociated by recent star
 formation \citep{Allen2002,Louie2013}.

 The observed velocity gradients in the position--velocity
 distribution of CO and [C\,{\sc ii}] are suggestive of the presence
 of galactic shocks that agglomerate molecular clouds (upstream,
 traced by CO) and trigger star formation (downstream, traced by
 [C\,{\sc ii}]).   The [C\,{\sc ii}] emission downstream from CO
   is likely associated with embedded star formation arising from
   dense ionized gas and/or the FUV illuminated surfaces of molecular
   clouds (PDRs). Because of the sensitivity of the [C\,{\sc ii}]
   intensity to volume density ($\propto n^2$;
   e.g. \citealt{Goldsmith2012}), the contribution from diffuse
   H\,{\sc i} gas to the [C\,{\sc ii}] intensity is likely negligible
   in these regions.

 \subsubsection{Comparison with Theoretical Models}
 \label{sec:comp-with-theor}

 \begin{figure}[t]
   \includegraphics[width=0.475\textwidth,angle=0]{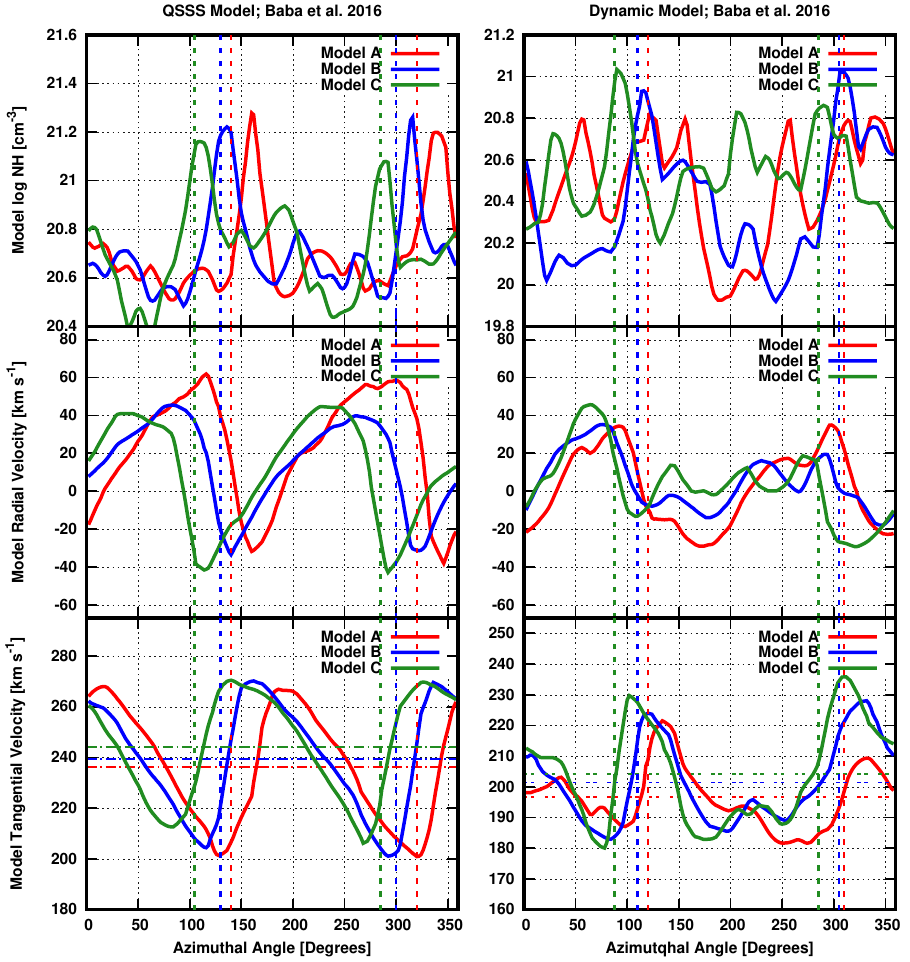}
\centering
\caption{Azimuthal distribution of the column density ({\it top
    panel}), radial velocity ({\it middle panel}), and tangential
  velocities ({\it bottom panel}) resulting from the QSSS and
  dynamical models presented by \citet[][ see also their Figure\,4 for
    further details]{Baba2016}. We also show in the location of the
  spiral potential and the average tangential velocities as vertical
  and horizontal lines, respectively. }\label{fig:velocity_theory}
\end{figure}

\begin{figure*}[t]
\includegraphics[width=0.75\textwidth,angle=0]{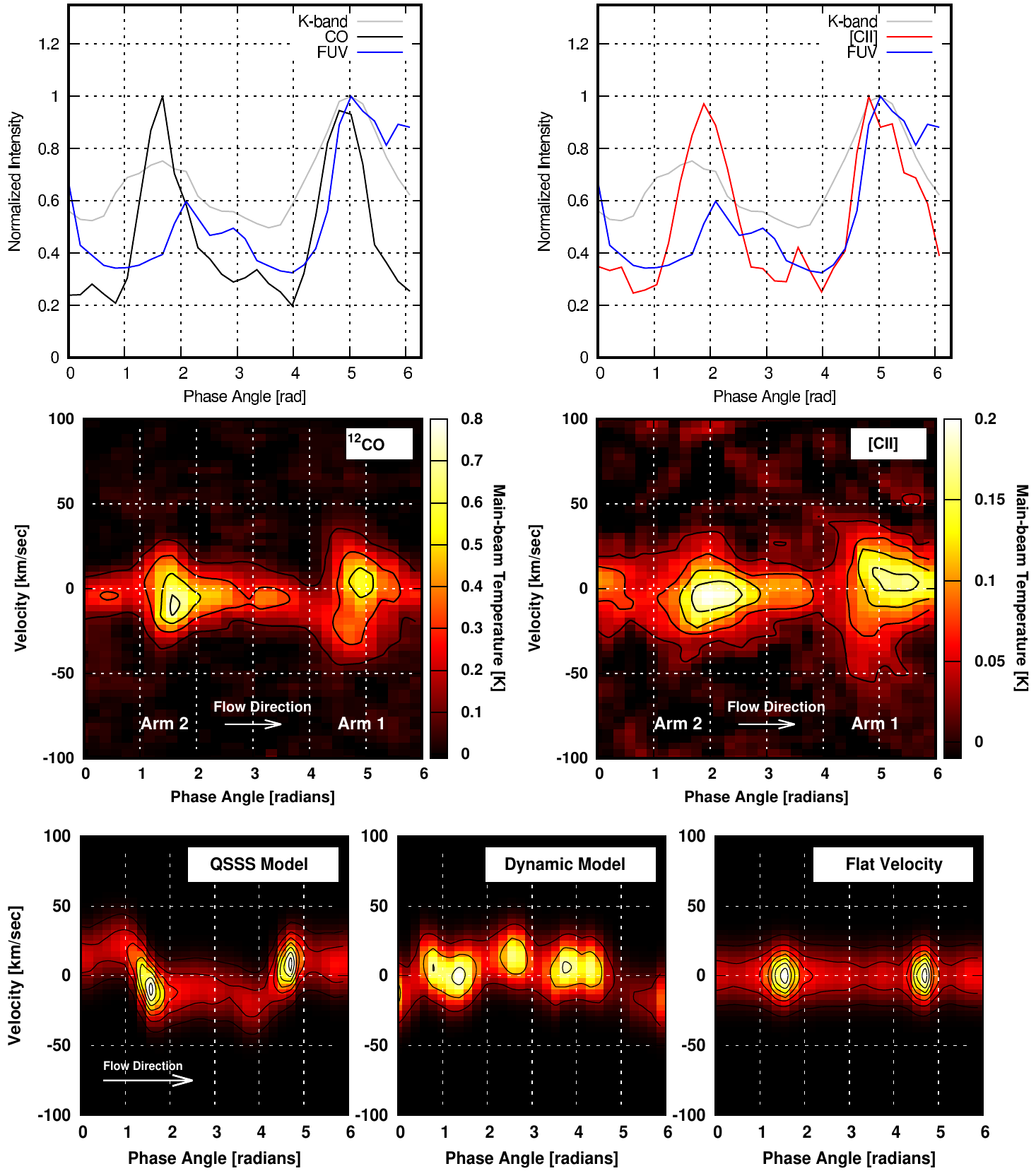}
\centering
\caption{{(\it Upper panels)} Integrated intensity distribution of
  $^{12}$CO ({\it left}) and [C\,{\sc ii}] ({\it right}) as a function
  of azimuthal angle between 2.6\,kpc and 3.7\,kpc from M51's
  center. We also include the azimuthal distribution of FUV and NIR
  K--band emission in both panels.  ({\it Middle panels})
  Position--velocity map of the $^{12}$CO ({\it left}) and [C\,{\sc
      ii}] ({\it right}) emission across the spiral arms between
  2.6\,kpc and 3.7\,kpc from M51's center. ({\it Bottom panels})
  Simulated position--velocity maps predicted from the QSSS ({\it
    left}) and dynamical  ({\it middle}) models by \citet{Baba2016}. We also include a purely
  circular rotation model ({\it right}).  The flow direction goes from
  left to right, as indicated by the white
  arrow.}\label{fig:sample_pv_1}
\end{figure*}
\begin{figure*}[t]
\includegraphics[width=0.75\textwidth,angle=0]{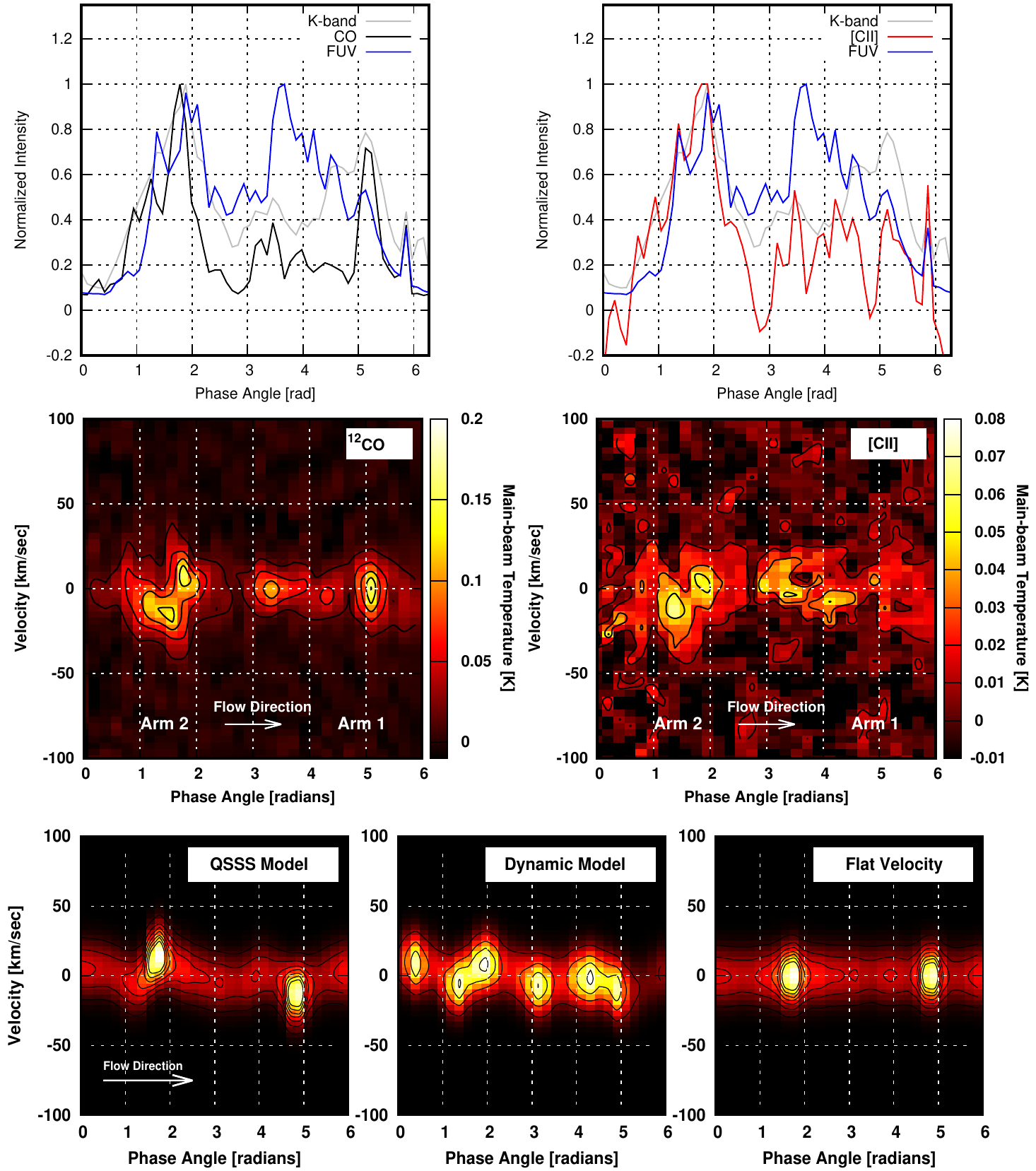}
\centering
\caption{{(\it Upper panels)} Integrated intensity distribution of
  $^{12}$CO ({\it left}) and [C\,{\sc ii}] ({\it right}) as a function
  of azimuthal angle between 6.9\,kpc and 7.9\,kpc from M51's
  center. We also include the azimuthal distribution of FUV and NIR
  K--band emission in both panels.  ({\it Middle panels})
  Position--velocity map of the $^{12}$CO ({\it left}) and [C\,{\sc
      ii}] ({\it right}) emission across the spiral arms between
  6.9\,kpc and 7.9\,kpc from M51's center. ({\it Bottom panels})
  Simulated position--velocity maps predicted from the QSSS ({\it
    left}) and dynamical ({\it middle}) models by \citet{Baba2016}. We
  also include a purely circular rotation model ({\it right}).  The
  flow direction goes from left to right, as indicated by the white
  arrow.   }\label{fig:sample_pv_2}
\end{figure*}

 We investigate whether the observed position velocity distribution of
 CO and [C\,{\sc ii}] is consistent with the existence of galactic
 shocks by comparing our observed position--velocity distributions
 across spiral arms with those predicted by theoretical models of the
 nature of spiral arms in galaxies. \citet{Baba2016} studied velocity
 patterns in hydrodynamical simulations of galaxies with QSSS and
 dynamical spirals arms. These simulations include self--gravity,
 radiative cooling, heating due to the interstellar FUV radiation
 field, and include a sub--grid model for star formation and stellar
 feedback.  They presented azimuthal distributions of column density,
 tangential and radial velocity profiles for three different distances
 to the center of the model galaxies, that are well within the
 co-rotation radius ($R<0.5 R_{\rm cr}$), and are separated by
 1\,kpc. We denote these three models, in order of increasing distance
 to the model galaxy's center, as Model A, B, and C. In
 Figure~\ref{fig:velocity_theory}, we show the azimuthal distribution
 of the column density, radial velocity, and tangential velocities
 resulting from the QSSS and dynamical models presented by \citet[][
   see their Figure\,4 for further details]{Baba2016}.  We also show
 in Figure~\ref{fig:velocity_theory}, the location of the spiral
 potential and the average tangential velocities as vertical and
 horizontal lines, respectively.  Spiral arms are manifested as peaks
 in the column density distribution, with the QSSS model showing gas
 peaks downstream from the spiral potential peak with offsets that
 increase with the distance to the galactic center, as expected for
 this theory inside the co--rotation radius, while in the dynamic
 model no systematic offset is seen. The shape of the radial and
 tangential profiles in both models are unaffected by different
 galactocentric distances.  In the QSSS model we see a periodic
 azimuthal distribution of both radial and tangential velocities, with
 spiral arms being associated with the radial velocity minimum and the
 average tangential velocity. In contrast, the dynamical model the
 spiral arms are associated with zero radial velocities and tangential
 velocities that are higher than their average.

\begin{figure*}[t]
\includegraphics[width=0.75\textwidth,angle=0]{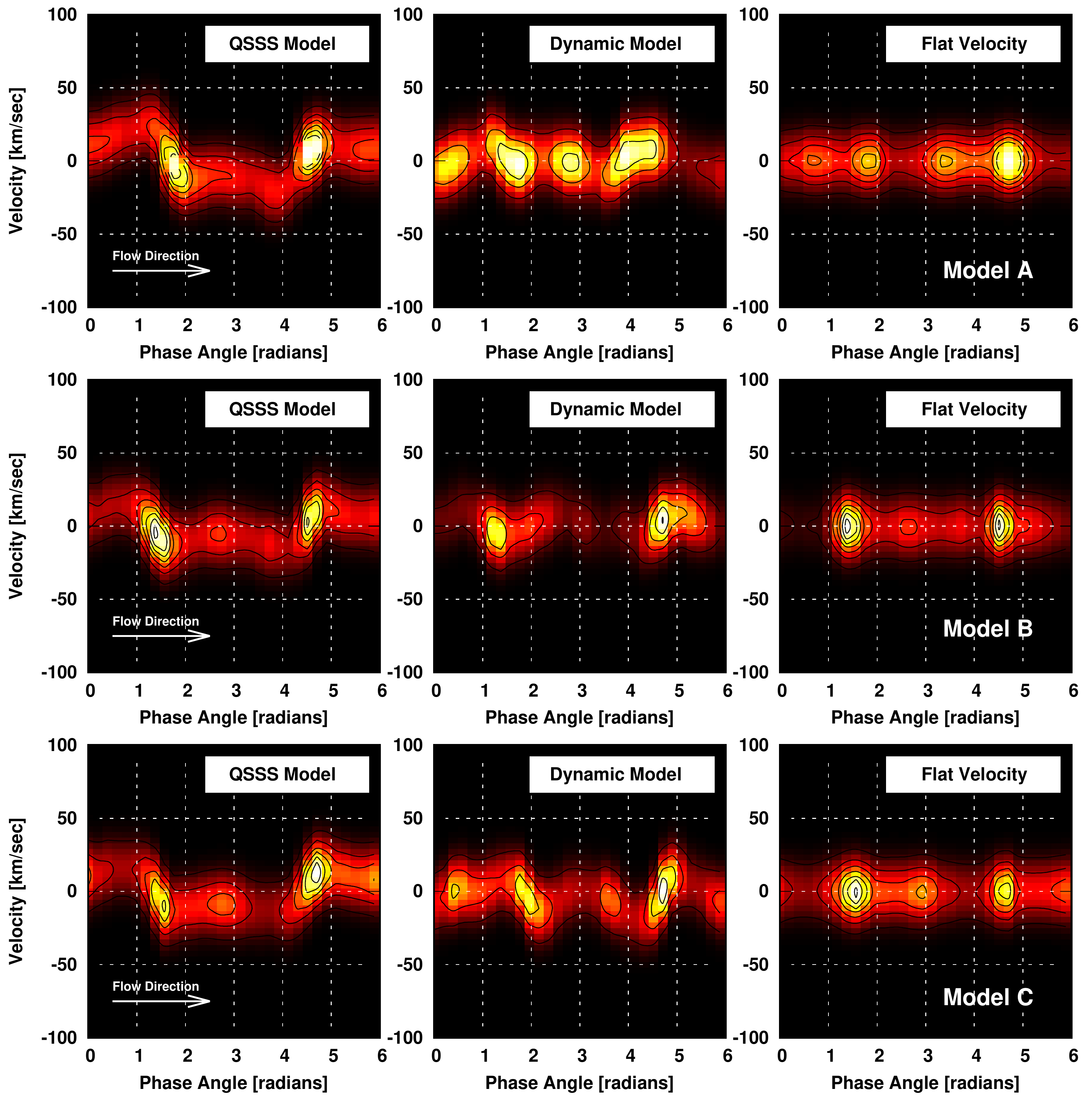}
\centering
\caption{Projected position--velocity maps of model galaxies with
  quasi--static (QSSS model) and dynamic spiral (Dynamic Model) arms
  presented by \citet{Baba2016} for Model A, B, and C
  (Section~\ref{sec:comp-with-theor}).  We also include a galaxy with
  no velocity structure other than pure rotation (Flat Velocity). The
  predicted velocities are projected to correspond to that observed at
  a ring between 2.6\,kpc and 3.7\,kpc from the center of M51 (see
  Figure~\ref{fig:sample_pv_1}).  }\label{fig:pvmodel_678kpc}
\end{figure*}

\begin{figure*}[t]
\includegraphics[width=0.75\textwidth,angle=0]{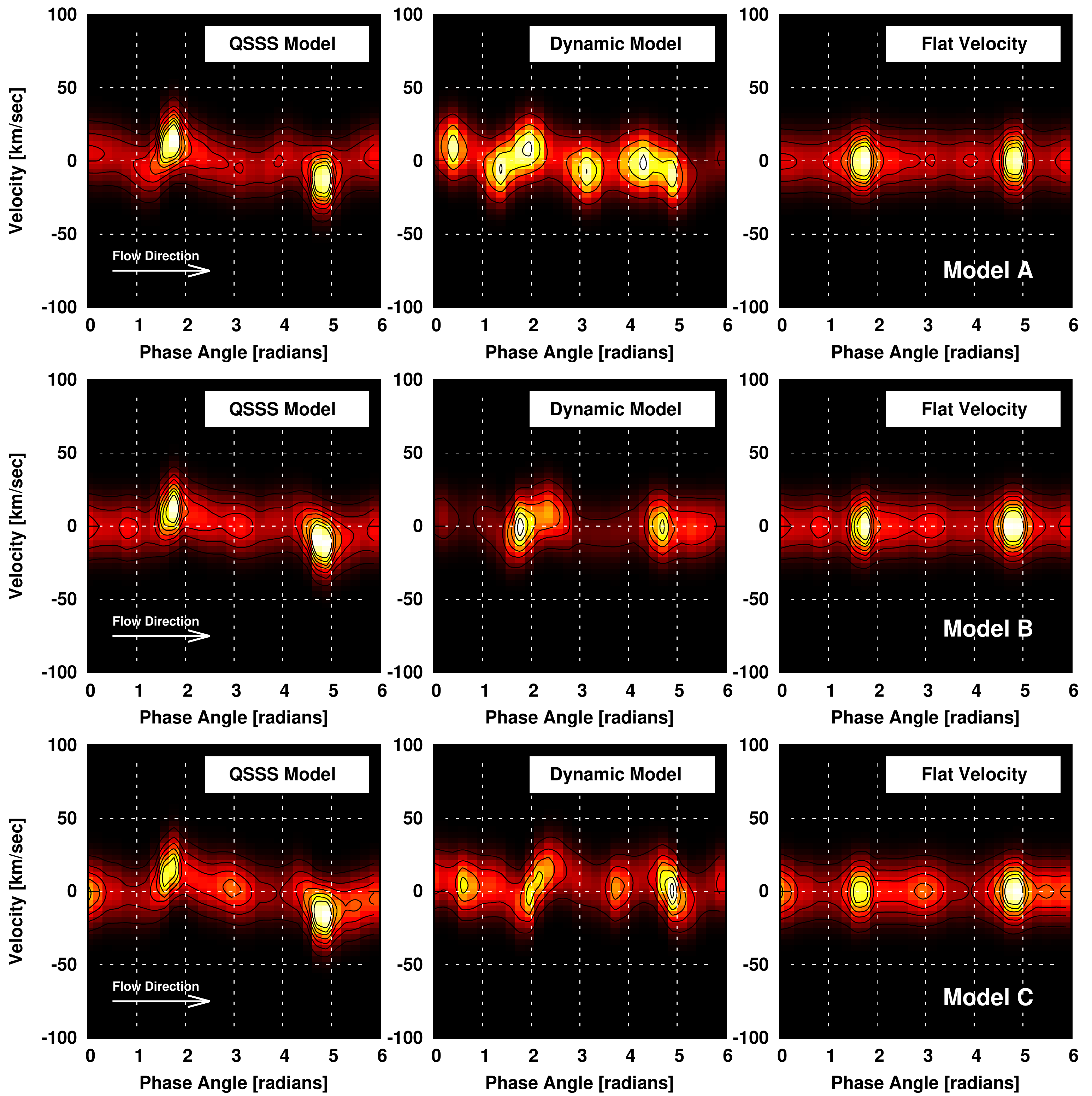}
\centering
\caption{Projected position--velocity maps of model galaxies with
  quasi--static (QSSS model) and dynamic spiral (Dynamic Model) arms
  presented by \citet{Baba2016} for Model A, B, and C
  (Section~\ref{sec:comp-with-theor}).  We also include a galaxy with
  no velocity structure other than pure rotation (Flat Velocity).  The
  predicted velocities are projected to correspond to that observed at
  a ring between 6.9\,kpc and 7.9\,kpc from the center of M51 (see
  Figure~\ref{fig:sample_pv_2}). }\label{fig:pvmodel_678kpv_outergal}
\end{figure*}

  We compare our observations with the predictions from
  \citet{Baba2016} by creating model galaxies with projected velocity
  structures that follows the radial and tangential velocity profiles
  of dynamical and steady models as shown in
  Figure~\ref{fig:velocity_theory}.  We also generate a purely
  circular model, i.e. with no velocity structure other than rotation,
  with the aim of determining whether any systematics in our method
  affects the shape of the position velocity distribution in spiral
  arms. The model galaxies are assumed to be at the same distance to
  M51 and have the same inclination, $i$, position angle, $\theta_{\it
    MA}$, systemic velocity of the galaxy, $V_{\rm sys}$
  (Table\,1). We calculate the projected velocities using
\begin{equation}\label{eq:3}
V=V_{sys}+[V_{\it R} \sin( \theta - \theta _{\it MA}) + V_{\it T}
\cos( \theta- \theta_{\it MA}) ]\sin(i),
\end{equation}
where $V_{\it R}$ and $V_{\it T}$ are the predicted radial and
tangential velocities (from Figure~\ref{fig:velocity_theory}),
respectively.  For each pixel we assumed Gaussian line profiles with a
FWHM line width of 30\,km\,s$^{-1}$ (typical of our [C\,{\sc ii}]
data) and peak intensities given by $T_{\rm peak}=N_{\rm H}/{\rm
  FWHM}$, where $N_{\rm H}$ is the hydrogen column density predicted
by the models shown in Figure~\ref{fig:velocity_theory}.  We generate
two sets of simulated galaxies with spiral arms with pitch angles that
correspond to those in M51's inner and outer galaxy (Table\,2).  To
correct the projected model velocities to a reference frame that
rotates with the spiral potential, we also produce a map of projected
velocities in which $V_{\it R}=0$\,km\,s$^{-1}$ and $V_{\it T}$ is
equal to the rotation velocity of the galaxy. The tangential velocity
shown in Figure~\ref{fig:velocity_theory} can be decomposed into the
rotation velocity plus peculiar velocities due to the spiral arm
perturbation. As we did with the M51 data, we set the velocity axis of
the simulated data cubes to be $V=0$\,km\,s$^{-1}$ at the projected
velocity of the purely rotating map. We show the position--velocity
resulting from QSSS and Dynamic Model A, together with observations,
in Figure~\ref{fig:sample_pv_1} and \ref{fig:sample_pv_2} for rings in
the inner and outer galaxy, respectively.  In
Figure~\ref{fig:pvmodel_678kpc} and \ref{fig:pvmodel_678kpv_outergal},
we also illustrate the predicted position velocity maps for Model A,
B, and C, for the inner and outer galaxy rings shown in
Figure~\ref{fig:sample_pv_1} and \ref{fig:sample_pv_2},
respectively. In Appendix~\ref{sec:depend-posit-veloc} we study the
sensitivity of the QSSS predicted velocity patterns on the amplitude
of radial and tangential motions and on the spatial location of spiral
arms in the galaxy.

In Figure~\ref{fig:sample_pv_1}, we show the position--velocity
distribution of [C\,{\sc ii}] and CO across spiral arms between
2.6\,kpc and 3.7\,kpc from M51's center. We also include the
distribution of [C\,{\sc ii}] and $^{12}$CO integrated intensities as
a function of phase angle and predictions for this region simulated
from Model A of \citet{Baba2016} and for a model galaxy with no
peculiar velocity other than its rotation\footnote{Note that the line
  width is constant for the position velocity map of the model galaxy
  with no peculiar velocity other than its rotation. The apparent
  variation of the line width distribution with phase angle seen in
  this position--velocity map is the result of the truncation in the
  color scale with the width at the truncation intensity varying from
  faint to bright regions.}.  We also include the NIR K--band
distribution, denoting the location of the bottom of the gravitational
potential, and FUV emission tracing evolved star formation regions.
The observed position velocity maps show an antisymmetric pattern with
a velocity gradient observed in both tracers. Following the flow
direction from left to right, the gas velocity increases and then
decreases in the first (left) spiral arm and later the velocity is
reduced and then increased at the location of the second spiral arm
(right).  The CO intensity peaks at the location of the second
velocity gradient in both arms. The [C\,{\sc ii}] intensity peaks
downstream from the CO peak in both spiral arms. Note that in the
integrated intensity distribution for this ring, we see no noticeable
offset between $^{12}$CO and [C\,{\sc ii}] for Arm\,1, while there is
a clear offset in the position velocity maps.  This discrepancy is due
to the variation of the line--widths as a function of phase angle,
which results in a broader integrated intensity distribution across
the arm, which makes it difficult for offsets to be identified in the
position-position space.
 As we can see, the QSSS model shows also an antisymmetric pattern in
 which the velocity increases and then decreases in the first spiral
 arm and then decreases and subsequently increases at the location of
 the second spiral arm. For the dynamical model we see variations in
 the velocity field but we do not see the antisymmetric pattern that
 is observed and predicted by the QSSS model. {\it Further evidence
   supporting the QSSS model is the fact that we see a displacement in
   position velocity space between CO and [C\,{\sc ii}], with the
   latter, which traces star formation, being downstream from the
   dense molecular gas traced by CO.} This result is a prediction of
 the QSSS model that is not seen in the dynamical model. In the latter
 model, the gas is expected to flow from {\it both} sides of the arm,
 and thus no relative displacement in these tracers are expected.  We
 therefore conclude that the spiral structure observed in this area in
 M51 is consistent with the QSSS spiral model.  Note that what we
 observe here is the shock and gas phase changes across the spiral
 arms, and therefore, the events that occur over arm--crossing
 timescales. Therefore, we cannot exclude a possibility that the
 "QSSS" spiral arms might move and evolve over a much longer
 timescale.

The projected position--velocity maps derived from Model A of
\citet{Baba2016}, shown in the bottom panels in
Figure~\ref{fig:sample_pv_2}, also show velocity gradient that is
opposite to that seen in the inner galaxy, due to the different
direction of the flow with respect to the spiral arms. For Arm 2, we
find a resemblance between the observations and predictions from the
QSSS theory. We observe a velocity gradient for the predictions for
the dynamic spiral at this radii, but the magnitude of the velocity
variation is smaller than that observed.  Additionally, this velocity
variation is not present for predictions in the other models from
\citet{Baba2016}, as shown in
Figure~\ref{fig:pvmodel_678kpv_outergal}.  The offsets between
$^{12}$CO and [C\,{\sc ii}] again suggest that Arm 2 is consistent
with the predictions from the QSSS theory.  However, the lack of
velocity structure in Arm 1 would also allow for the dynamic spiral
mechanism to govern the structure of this spiral arm in the outer
regions of M51.  This spiral arm is pointing away from the
  companion in the outer galaxy, and therefore is feeling less of the
  direct tidal pull, perhaps making it more flocculent in nature.

 Note that our comparison with the \citet{Baba2016} calculations
  is qualitative, as their model parameters do not necessarily match
  those of the M51 galaxy. The QSSS is triggered when the rotation
  speed of a spiral potential is slower/faster than that of gas, and
  the dynamical spiral arm is governed primarily by the swing
  amplification due to epicyclic motions. These two physical
  mechanisms can operate in any spiral galaxy, and hence the two
  models are expected to make qualitative differences independent of
  the total stellar and gas masses of a galaxy and its rotation curve.
  In particular, their method was developed for isolated galaxies,
  while M51 is affected by the interaction with the M51b galaxy
  \citep{Pettitt2017,Tress2019}.  The companion galaxy may be causing
  a significant $m=2$ mode perturbation, which may make M51 a special
  case, and there are suggestions that in tidally interacting systems
  shocks similar to those predicted by the QSSS theory could be
  present \citep{Oh2008,Dobbs2010,Oh2015}.  To any extent, this is a
  case study, and we need a larger sample of galaxies to draw a
  general conclusion on the mechanism for driving spiral arms.


\subsubsection{Comparison with previous studies of the nature of spiral arms in M51}

 Another test for distinguishing between different theories of the
 nature of spiral arms in galaxies is the distribution of star
 clusters age across spiral arms.  \citet{Dobbs2010b} presented
 numerical simulations for galaxies with QSSS, dynamic spirals, a
 barred galaxy, and an interacting galaxies like M51.  They find that
 in the case of the QSSS and barred--galaxy, stellar clusters are
 predicted to show offsets from the spiral arms that increase with
 age, as stars and gas rotate faster than the spiral pattern, and
 stars formed after the compression of gas in the spiral arms
 eventually overtake the spiral pattern as they age.  In the case of
 the dynamic model, stars form as a result of local gas instabilities,
 and thus no age gradient is expected.  They also predicted that for
 an interacting system, no clear stellar gradient is expected, due to
 the complex dynamics of the interaction.  These theoretical
 predictions have been tested with optical observations in
 M51. \citet{Chandar2017} used an optically derived catalog of star
 cluster ages to study their distribution across M51's spiral arms.
 They observe that young stellar clusters ($<$10\,Myr) are
 preferentially observed near the arms, but intermediate age
 (10--50\,Myr) and old (50--100\,Myr) have a more spread distribution.
 While they find an offset between CO emission and young ($<$10\,Myr)
 stellar clusters, which they interpreted as to be in agreement with
 QSSS spirals, they do not find a significant offsets from the spiral
 arms for older stellar clusters. \citet{Shabani2018} used a similar
 data set in M51 and also found no significant offsets in the location
 of stellar clusters across the arms of M51, suggesting that the
 dynamic model would also explain the nature of spiral arms in M51.
 Note, however, that for a rotation velocity of
 $\sim$200\,km\,s$^{-1}$ \citep{Meidt2013,Oikawa2014}, in the inner
 M51 ($R=3.2$\,kpc; see Figure~\ref{fig:sample_pv_1}) it would take a
 newly formed cluster only $\sim$40\,Myr to move from one arm to
 another, and therefore the young clusters observed near spiral arms
 are likely mixed with older clusters formed in the other spiral arm,
 or in the inter--arm regions, where CO and [C\,{\sc ii}] emission is
 also detected.  The rapid spatial mixing in the inner M51 makes the
 observation of a cluster age gradient difficult and therefore creates
 an ambiguity on whether stellar cluster age observations can be
 explained with the QSSS or dynamic model. The observed CO velocity
 pattern and offset between CO and [C\,{\sc ii}] provides an
 alternative evidence that QSSS is producing the spiral arms in M51.

  Based on the rotation velocity mentioned above, and the observed
  offsets in phase angle, it would take the gas about $\sim$5\,Myr to
  move between the CO and [C\,{\sc ii}] peaks in the
  position--velocity maps. The [C\,{\sc ii}] peak in the
  position--velocity maps is likely associated with the embedded star
  formation phase \citep[lasting for about 1.5\,Myr;
    e.g.][]{Kruijssen2019}, as it is located downstream from CO but
  upstream from FUV emission peaks, tracing the dense gas phase prior
  to star formation and the cloud dispersal by stellar feedback phase,
  respectively. Therefore, the observed offset between CO and [C\,{\sc
      ii}] peaks in position--velocity space suggest a star formation
  timescale of about 5\,Myr.


\section{Summary}
\label{sec:conclusions}

In this paper, we studied the influence of spiral density waves on the
evolution of the interstellar medium and star formation in the M51
galaxy.  We used new spectrally resolved upGREAT/SOFIA [C\,{\sc ii}]
data combined with NIR K--band, FUV, H\,{\sc i}, and CO data to study
the spatial and spatial--velocity distribution of different ISM phases
in the spiral arms of M51. Our results can be summarized as follows:

\begin{itemize}

\item We identify azimuthal offsets between NIR K--band, $^{12}$CO,
  [C\,{\sc ii}], and FUV, tracing stellar mass, dense and cold
  molecular gas, obscured star formation, and unobscured star
  formation, respectively, in the spiral arms of M51. However, we
  could not find systematic variations of these offsets with
  galactocentric distance at the angular resolution of our
  observations.
     
   \item The offsets between $^{12}$CO and [C\,{\sc ii}] in M51 are
     more apparent in Arm\,2, connecting to the companion galaxy M51b,
     compared with Arm\,1, pointing away from the companion. We find
     that identifying offsets between $^{12}$CO and [C\,{\sc ii}]
     integrated intensities is complicated by the varying line widths
     of these tracers at the location of spiral arms, and that they
     are better separated when comparing peak main beam temperatures
     in the position velocity space.

   \item The position velocity maps of H\,{\sc i}, $^{12}$CO, and
     [C\,{\sc ii}] across spiral arms in M51 show strong velocity
     gradients at the location of stellar arms (traced by K--band
     data) with a clear offset in position velocity space between
     upstream molecular gas (traced by $^{12}$CO) and downstream star
     formation (traced by [C\,{\sc ii}]).

  \item We compared the observed position velocity maps across spiral
    arms with simulated observations from numerical simulations of
    galaxies with both dynamical and quasi--stationary steady spiral
    arms that predict tangential and radial velocities at the location
    of spiral arms.  We find that our observations are consistent with
    the presence of spiral shock in spiral arms in the inner M51 and
    in the arm connecting to the companion galaxy, M51b, in the outer
    M51 (Arm 2).

\end{itemize}

  Our analysis shows that spectrally resolved observations are
  important tools for studying kinematics of spiral arms and the
  evolution of the interstellar medium in galaxies. They are also
  useful for distinguishing between competing theories of the nature
  of spiral structure in galaxies. We speculate that the spiral shocks
  observed in M51 might originate as a result of the interaction
  between M51 and M51b. A better test for theories of spiral structure
  in isolated galaxies would be to apply the techniques described in
  this paper to velocity resolved [C\,{\sc ii}] and CO observations of
  systems that have not been influenced by tidal interaction. Both
  [C\,{\sc ii}] and CO are needed to distinguish between steady and
  dynamic spirals, as the velocity gradients predicted for steady
  spirals can be mimicked by dynamic spirals. However, the observed
  offsets between [C\,{\sc ii}] and CO, which are often only seen in
  position velocity space and are predicted only by the QSSS theory,
  were able to break the ambiguity between these models.


\begin{acknowledgements}

  We thank an anonymous referee for helpful comments on the
  manuscript.  We also thank Drs. William D. Langer and Youngmin Seo
  for useful comments on the manuscript.  This research was conducted
  in part at the Jet Propulsion Laboratory, California Institute of
  Technology under contract with the National Aeronautics and Space
  Administration (80NM0018D0004). The development of upGREAT was
  financed by the participating institutes, by the Federal Ministry of
  Economics and Technology via the German Space Agency (DLR) under
  Grants 50 OK 1102, 50 OK 1103 and 50 OK 1104 and within the
  Collaborative Research Centre 956, sub-projects D2 and D3, funded by
  the Deutsche Forschungsgemeinschaft (DFG). RSK and SCOG receive
  financial support from the Deutsche Forschungsgemeinschaft (DFG,
  German Research Foundation) -- Project-ID 138713538 -- SFB 881
  (``The Milky Way System'', subprojects A1, B1, B2, B8), and they
  acknowledges support fom the DFG via the Heidelberg Cluster of
  Excellence {\em STRUCTURES} in the framework of Germany’s Excellence
  Strategy (grant EXC-2181/1 - 390900948).  We thank the staff of the
  SOFIA Science Center for their help.  \copyright\ 2020. All rights
  reserved.

\end{acknowledgements}

\newpage
\appendix

\section{Dependence of position--velocity patterns on tangential and radial velocity amplitudes, and location in the galaxy.}
\label{sec:depend-posit-veloc}

\begin{figure*}[h]
\includegraphics[width=0.95\textwidth,angle=0]{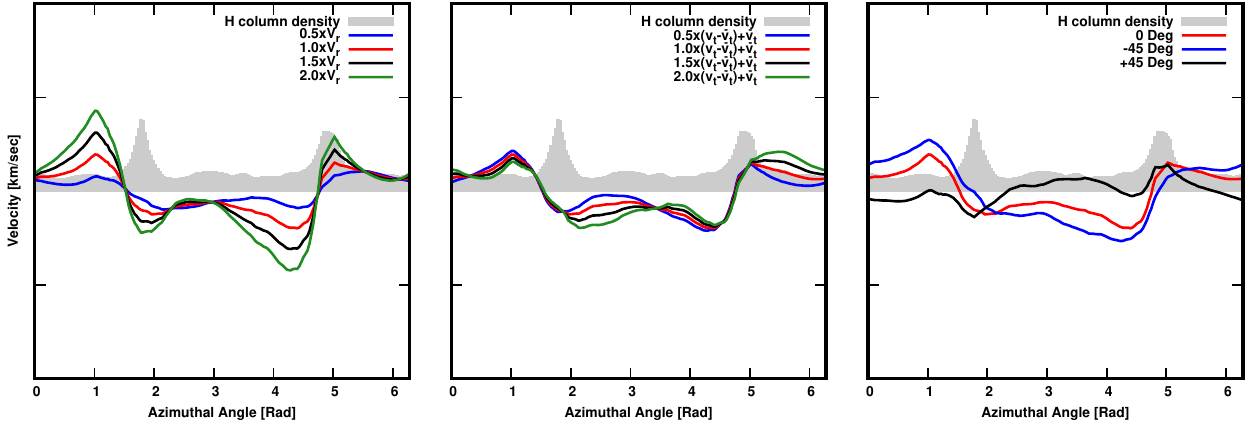}
\centering
\caption{({\it Left panel}) Projected velocity as a function of phase
  angle derived from the QSSS Model A in \citet{Baba2016}, with the radial
  velocity contribution being factored by factors of 0.5, 1, 1.5, and
  2, and the tangential velocity contribution kept as predicted by the
  model. ({\it Middle panel}) Projected velocity as a function of
  phase angle taken from QSSS Model A, with the tangential velocity
  contribution multiplied by factors of 0.5, 1, 1.5, and 2, and the
  radial velocity contribution kept as predicted by the model. ({\it
    Right panel}) Projected velocity as a function of phase angle,
  again taken from QSSS Model A, but with the angular location being
  rotated by $-$45 and $+$45 degrees from the location of the spiral arms
  for a radius between 2.6\, and 3.7\,kpc in our simulated M51
  galaxy. In all panels we also show the hydrogen column density
  distribution as a function of phase angle from QSSS Model A to
  denote the location of spiral arms peaks.
}\label{fig:amplification_dependence}
\end{figure*}

 The similarity between the observed and QSSS model--predicted
 velocity distribution motivate us to use model--predicted radial and
 tangential velocities to investigate under what conditions the
 observed velocity profiles are produced. In the left panel of
 Figure~\ref{fig:amplification_dependence}, we show the position peak
 velocity maps for \citet{Baba2016} Model A, where we multiplied the
 radial velocity by factors of 0.5, 1, 1.5, and 2, while keeping the
 tangential velocity distribution as predicted by the model. The
 projected velocity is determined using Equation~\ref{eq:1} and
 subtracting the projected velocity in the case when only the galaxy
 rotation is present.  Increasing the amplitude of the radial velocity
 results in larger velocity gradients in the position velocity
 maps. As discussed in Section~\ref{sec:comp-with-theor} above, in the
 QSSS model the spiral arms are associated with V$_r$ minima and V$_t$
 close to its average, and thus the contribution from the radial
 velocity to the projected velocity is maximized at the location of
 spiral arms.  In the middle panel of
 Figure~\ref{fig:amplification_dependence}, we show the
 position--velocity maps for Model A in the case where the tangential
 velocity is multiplied by factors of 0.5, 1, 1.5, and 2, while
 keeping the radial velocity distribution as predicted by the
 model. Note that in this case we vary the amplitude of the tangential
 velocity with respect to its average value, as it also includes the
 rotation of the galaxy. No significant change in the position
 velocity maps is observed in this case.

Another important condition for QSSS velocity patterns to be
observable is the azimuthal angle of spiral arms for a given radius.
In the inner M51, the azimuthal angles of the spiral arms where we see
a velocity gradient (between 2.6\,kpc and 3.7\,kpc;
Figure~\ref{fig:sample_pv_1}) are $\theta\simeq 124\degr$ and
$34\degr$ for Arm 1 and Arm 2, respectively. These azimuthal angles
correspond to about $\Delta \theta=+45$\degr\ and $-$45\degr\ offsets
from the minor axis for Arm 1 and Arm 2, respectively. In the outer
galaxy location where we see a velocity gradient (between 6.9\,kpc and
7.9\,kpc; Figure~\ref{fig:sample_pv_2}) the azimuthal angle of Arm 2
is 158\degr, which is about $\Delta \theta \simeq -11.7$\degr\ from
the minor axis.  In the right panel of
Figure~\ref{fig:amplification_dependence}, we show the predicted
position velocity profiles for the QSSS Model A galaxy with the same
azimuthal angle as that observed in the inner M51. We also show the
position--velocity distribution for spiral arms located being rotated
by $\Delta \theta=-$45\degr (i.e located at the minor axis) and
$+45$\degr (i.e. located 90\degr\ from the minor axis). Note that we
moved the origin of the phase angle distribution by the same azimuthal
angles shown in the figure to keep the location of spiral arms in
phase angle constant.  We see that for an azimuthal angle of the arms
that are at the location of the minor axis (offset by $-$45\degr) the
projected velocity gradient shows the largest amplitude, but it
dissapears when we rotate the spiral ams by an offset of $+$45\degr\ from
the location where we see the arms in the inner M51.  From
Equation~(\ref{eq:3}) we see that the contribution from the radial
velocity is maximized, and the contribution of the tangential velocity
is minimized, at locations close to the minor axes
($\sin(\theta-\theta_{\rm MA})\simeq 1$). As shown above, the velocity
gradients seen in the projected velocity in the QSSS model mostly
depend on the radial velocity and thus the observed the pattern is
expected in rings where the arms are at an azimuthal angle that
maximizes the contribution from the radial velocity to the projected
velocity.
%
%

\bibliographystyle{aasjournal}

\bibliography{/home/jpineda/pCloudDrive/latex/papers}

\clearpage

\end{document}